\documentclass[useAMS,usenatbib, referee]{mn2e}
\usepackage{times}
\usepackage{url}
\usepackage{graphicx}
\usepackage{deluxetable}
\usepackage[usenames]{color}
\DeclareGraphicsExtensions{.pdf,.png,.jpg,.mps,.eps,.ps}
\usepackage{amsmath}
\usepackage{natbib}
\usepackage{soul}
\definecolor{AliceBlue}{rgb}{0.94,0.97,1.00}
\definecolor{AntiqueWhite1}{rgb}{1.00,0.94,0.86}
\definecolor{AntiqueWhite2}{rgb}{0.93,0.87,0.80}
\definecolor{AntiqueWhite3}{rgb}{0.80,0.75,0.69}
\definecolor{AntiqueWhite4}{rgb}{0.55,0.51,0.47}
\definecolor{AntiqueWhite}{rgb}{0.98,0.92,0.84}
\definecolor{BlanchedAlmond}{rgb}{1.00,0.92,0.80}
\definecolor{BlueViolet}{rgb}{0.54,0.17,0.89}
\definecolor{CadetBlue1}{rgb}{0.60,0.96,1.00}
\definecolor{CadetBlue2}{rgb}{0.56,0.90,0.93}
\definecolor{CadetBlue3}{rgb}{0.48,0.77,0.80}
\definecolor{CadetBlue4}{rgb}{0.33,0.53,0.55}
\definecolor{CadetBlue}{rgb}{0.37,0.62,0.63}
\definecolor{CornflowerBlue}{rgb}{0.39,0.58,0.93}
\definecolor{DarkBlue}{rgb}{0.00,0.00,0.55}
\definecolor{DarkCyan}{rgb}{0.00,0.55,0.55}
\definecolor{DarkGoldenrod1}{rgb}{1.00,0.73,0.06}
\definecolor{DarkGoldenrod2}{rgb}{0.93,0.68,0.05}
\definecolor{DarkGoldenrod3}{rgb}{0.80,0.58,0.05}
\definecolor{DarkGoldenrod4}{rgb}{0.55,0.40,0.03}
\definecolor{DarkGoldenrod}{rgb}{0.72,0.53,0.04}
\definecolor{DarkGray}{rgb}{0.66,0.66,0.66}
\definecolor{DarkGreen}{rgb}{0.00,0.39,0.00}
\definecolor{DarkGrey}{rgb}{0.66,0.66,0.66}
\definecolor{DarkKhaki}{rgb}{0.74,0.72,0.42}
\definecolor{DarkMagenta}{rgb}{0.55,0.00,0.55}
\definecolor{DarkOliveGreen1}{rgb}{0.79,1.00,0.44}
\definecolor{DarkOliveGreen2}{rgb}{0.74,0.93,0.41}
\definecolor{DarkOliveGreen3}{rgb}{0.64,0.80,0.35}
\definecolor{DarkOliveGreen4}{rgb}{0.43,0.55,0.24}
\definecolor{DarkOliveGreen}{rgb}{0.33,0.42,0.18}
\definecolor{DarkOrange1}{rgb}{1.00,0.50,0.00}
\definecolor{DarkOrange2}{rgb}{0.93,0.46,0.00}
\definecolor{DarkOrange3}{rgb}{0.80,0.40,0.00}
\definecolor{DarkOrange4}{rgb}{0.55,0.27,0.00}
\definecolor{DarkOrange}{rgb}{1.00,0.55,0.00}
\definecolor{DarkOrchid1}{rgb}{0.75,0.24,1.00}
\definecolor{DarkOrchid2}{rgb}{0.70,0.23,0.93}
\definecolor{DarkOrchid3}{rgb}{0.60,0.20,0.80}
\definecolor{DarkOrchid4}{rgb}{0.41,0.13,0.55}
\definecolor{DarkOrchid}{rgb}{0.60,0.20,0.80}
\definecolor{DarkRed}{rgb}{0.55,0.00,0.00}
\definecolor{DarkSalmon}{rgb}{0.91,0.59,0.48}
\definecolor{DarkSeaGreen1}{rgb}{0.76,1.00,0.76}
\definecolor{DarkSeaGreen2}{rgb}{0.71,0.93,0.71}
\definecolor{DarkSeaGreen3}{rgb}{0.61,0.80,0.61}
\definecolor{DarkSeaGreen4}{rgb}{0.41,0.55,0.41}
\definecolor{DarkSeaGreen}{rgb}{0.56,0.74,0.56}
\definecolor{DarkSlateBlue}{rgb}{0.28,0.24,0.55}
\definecolor{DarkSlateGray1}{rgb}{0.59,1.00,1.00}
\definecolor{DarkSlateGray2}{rgb}{0.55,0.93,0.93}
\definecolor{DarkSlateGray3}{rgb}{0.47,0.80,0.80}
\definecolor{DarkSlateGray4}{rgb}{0.32,0.55,0.55}
\definecolor{DarkSlateGray}{rgb}{0.18,0.31,0.31}
\definecolor{DarkSlateGrey}{rgb}{0.18,0.31,0.31}
\definecolor{DarkTurquoise}{rgb}{0.00,0.81,0.82}
\definecolor{DarkViolet}{rgb}{0.58,0.00,0.83}
\definecolor{DeepPink1}{rgb}{1.00,0.08,0.58}
\definecolor{DeepPink2}{rgb}{0.93,0.07,0.54}
\definecolor{DeepPink3}{rgb}{0.80,0.06,0.46}
\definecolor{DeepPink4}{rgb}{0.55,0.04,0.31}
\definecolor{DeepPink}{rgb}{1.00,0.08,0.58}
\definecolor{DeepSkyBlue1}{rgb}{0.00,0.75,1.00}
\definecolor{DeepSkyBlue2}{rgb}{0.00,0.70,0.93}
\definecolor{DeepSkyBlue3}{rgb}{0.00,0.60,0.80}
\definecolor{DeepSkyBlue4}{rgb}{0.00,0.41,0.55}
\definecolor{DeepSkyBlue}{rgb}{0.00,0.75,1.00}
\definecolor{DimGray}{rgb}{0.41,0.41,0.41}
\definecolor{DimGrey}{rgb}{0.41,0.41,0.41}
\definecolor{DodgerBlue1}{rgb}{0.12,0.56,1.00}
\definecolor{DodgerBlue2}{rgb}{0.11,0.53,0.93}
\definecolor{DodgerBlue3}{rgb}{0.09,0.45,0.80}
\definecolor{DodgerBlue4}{rgb}{0.06,0.31,0.55}
\definecolor{DodgerBlue}{rgb}{0.12,0.56,1.00}
\definecolor{FloralWhite}{rgb}{1.00,0.98,0.94}
\definecolor{ForestGreen}{rgb}{0.13,0.55,0.13}
\definecolor{GhostWhite}{rgb}{0.97,0.97,1.00}
\definecolor{GreenYellow}{rgb}{0.68,1.00,0.18}
\definecolor{HotPink1}{rgb}{1.00,0.43,0.71}
\definecolor{HotPink2}{rgb}{0.93,0.42,0.65}
\definecolor{HotPink3}{rgb}{0.80,0.38,0.56}
\definecolor{HotPink4}{rgb}{0.55,0.23,0.38}
\definecolor{HotPink}{rgb}{1.00,0.41,0.71}
\definecolor{IndianRed1}{rgb}{1.00,0.42,0.42}
\definecolor{IndianRed2}{rgb}{0.93,0.39,0.39}
\definecolor{IndianRed3}{rgb}{0.80,0.33,0.33}
\definecolor{IndianRed4}{rgb}{0.55,0.23,0.23}
\definecolor{IndianRed}{rgb}{0.80,0.36,0.36}
\definecolor{LavenderBlush1}{rgb}{1.00,0.94,0.96}
\definecolor{LavenderBlush2}{rgb}{0.93,0.88,0.90}
\definecolor{LavenderBlush3}{rgb}{0.80,0.76,0.77}
\definecolor{LavenderBlush4}{rgb}{0.55,0.51,0.53}
\definecolor{LavenderBlush}{rgb}{1.00,0.94,0.96}
\definecolor{LawnGreen}{rgb}{0.49,0.99,0.00}
\definecolor{LemonChiffon1}{rgb}{1.00,0.98,0.80}
\definecolor{LemonChiffon2}{rgb}{0.93,0.91,0.75}
\definecolor{LemonChiffon3}{rgb}{0.80,0.79,0.65}
\definecolor{LemonChiffon4}{rgb}{0.55,0.54,0.44}
\definecolor{LemonChiffon}{rgb}{1.00,0.98,0.80}
\definecolor{LightBlue1}{rgb}{0.75,0.94,1.00}
\definecolor{LightBlue2}{rgb}{0.70,0.87,0.93}
\definecolor{LightBlue3}{rgb}{0.60,0.75,0.80}
\definecolor{LightBlue4}{rgb}{0.41,0.51,0.55}
\definecolor{LightBlue}{rgb}{0.68,0.85,0.90}
\definecolor{LightCoral}{rgb}{0.94,0.50,0.50}
\definecolor{LightCyan1}{rgb}{0.88,1.00,1.00}
\definecolor{LightCyan2}{rgb}{0.82,0.93,0.93}
\definecolor{LightCyan3}{rgb}{0.71,0.80,0.80}
\definecolor{LightCyan4}{rgb}{0.48,0.55,0.55}
\definecolor{LightCyan}{rgb}{0.88,1.00,1.00}
\definecolor{LightGoldenrod1}{rgb}{1.00,0.93,0.55}
\definecolor{LightGoldenrod2}{rgb}{0.93,0.86,0.51}
\definecolor{LightGoldenrod3}{rgb}{0.80,0.75,0.44}
\definecolor{LightGoldenrod4}{rgb}{0.55,0.51,0.30}
\definecolor{LightGoldenrodYellow}{rgb}{0.98,0.98,0.82}
\definecolor{LightGoldenrod}{rgb}{0.93,0.87,0.51}
\definecolor{LightGray}{rgb}{0.83,0.83,0.83}
\definecolor{LightGreen}{rgb}{0.56,0.93,0.56}
\definecolor{LightGrey}{rgb}{0.83,0.83,0.83}
\definecolor{LightPink1}{rgb}{1.00,0.68,0.73}
\definecolor{LightPink2}{rgb}{0.93,0.64,0.68}
\definecolor{LightPink3}{rgb}{0.80,0.55,0.58}
\definecolor{LightPink4}{rgb}{0.55,0.37,0.40}
\definecolor{LightPink}{rgb}{1.00,0.71,0.76}
\definecolor{LightSalmon1}{rgb}{1.00,0.63,0.48}
\definecolor{LightSalmon2}{rgb}{0.93,0.58,0.45}
\definecolor{LightSalmon3}{rgb}{0.80,0.51,0.38}
\definecolor{LightSalmon4}{rgb}{0.55,0.34,0.26}
\definecolor{LightSalmon}{rgb}{1.00,0.63,0.48}
\definecolor{LightSeaGreen}{rgb}{0.13,0.70,0.67}
\definecolor{LightSkyBlue1}{rgb}{0.69,0.89,1.00}
\definecolor{LightSkyBlue2}{rgb}{0.64,0.83,0.93}
\definecolor{LightSkyBlue3}{rgb}{0.55,0.71,0.80}
\definecolor{LightSkyBlue4}{rgb}{0.38,0.48,0.55}
\definecolor{LightSkyBlue}{rgb}{0.53,0.81,0.98}
\definecolor{LightSlateBlue}{rgb}{0.52,0.44,1.00}
\definecolor{LightSlateGray}{rgb}{0.47,0.53,0.60}
\definecolor{LightSlateGrey}{rgb}{0.47,0.53,0.60}
\definecolor{LightSteelBlue1}{rgb}{0.79,0.88,1.00}
\definecolor{LightSteelBlue2}{rgb}{0.74,0.82,0.93}
\definecolor{LightSteelBlue3}{rgb}{0.64,0.71,0.80}
\definecolor{LightSteelBlue4}{rgb}{0.43,0.48,0.55}
\definecolor{LightSteelBlue}{rgb}{0.69,0.77,0.87}
\definecolor{LightYellow1}{rgb}{1.00,1.00,0.88}
\definecolor{LightYellow2}{rgb}{0.93,0.93,0.82}
\definecolor{LightYellow3}{rgb}{0.80,0.80,0.71}
\definecolor{LightYellow4}{rgb}{0.55,0.55,0.48}
\definecolor{LightYellow}{rgb}{1.00,1.00,0.88}
\definecolor{LimeGreen}{rgb}{0.20,0.80,0.20}
\definecolor{MediumAquamarine}{rgb}{0.40,0.80,0.67}
\definecolor{MediumBlue}{rgb}{0.00,0.00,0.80}
\definecolor{MediumOrchid1}{rgb}{0.88,0.40,1.00}
\definecolor{MediumOrchid2}{rgb}{0.82,0.37,0.93}
\definecolor{MediumOrchid3}{rgb}{0.71,0.32,0.80}
\definecolor{MediumOrchid4}{rgb}{0.48,0.22,0.55}
\definecolor{MediumOrchid}{rgb}{0.73,0.33,0.83}
\definecolor{MediumPurple1}{rgb}{0.67,0.51,1.00}
\definecolor{MediumPurple2}{rgb}{0.62,0.47,0.93}
\definecolor{MediumPurple3}{rgb}{0.54,0.41,0.80}
\definecolor{MediumPurple4}{rgb}{0.36,0.28,0.55}
\definecolor{MediumPurple}{rgb}{0.58,0.44,0.86}
\definecolor{MediumSeaGreen}{rgb}{0.24,0.70,0.44}
\definecolor{MediumSlateBlue}{rgb}{0.48,0.41,0.93}
\definecolor{MediumSpringGreen}{rgb}{0.00,0.98,0.60}
\definecolor{MediumTurquoise}{rgb}{0.28,0.82,0.80}
\definecolor{MediumVioletRed}{rgb}{0.78,0.08,0.52}
\definecolor{MidnightBlue}{rgb}{0.10,0.10,0.44}
\definecolor{MintCream}{rgb}{0.96,1.00,0.98}
\definecolor{MistyRose1}{rgb}{1.00,0.89,0.88}
\definecolor{MistyRose2}{rgb}{0.93,0.84,0.82}
\definecolor{MistyRose3}{rgb}{0.80,0.72,0.71}
\definecolor{MistyRose4}{rgb}{0.55,0.49,0.48}
\definecolor{MistyRose}{rgb}{1.00,0.89,0.88}
\definecolor{NavajoWhite1}{rgb}{1.00,0.87,0.68}
\definecolor{NavajoWhite2}{rgb}{0.93,0.81,0.63}
\definecolor{NavajoWhite3}{rgb}{0.80,0.70,0.55}
\definecolor{NavajoWhite4}{rgb}{0.55,0.47,0.37}
\definecolor{NavajoWhite}{rgb}{1.00,0.87,0.68}
\definecolor{NavyBlue}{rgb}{0.00,0.00,0.50}
\definecolor{OldLace}{rgb}{0.99,0.96,0.90}
\definecolor{OliveDrab1}{rgb}{0.75,1.00,0.24}
\definecolor{OliveDrab2}{rgb}{0.70,0.93,0.23}
\definecolor{OliveDrab3}{rgb}{0.60,0.80,0.20}
\definecolor{OliveDrab4}{rgb}{0.41,0.55,0.13}
\definecolor{OliveDrab}{rgb}{0.42,0.56,0.14}
\definecolor{OrangeRed1}{rgb}{1.00,0.27,0.00}
\definecolor{OrangeRed2}{rgb}{0.93,0.25,0.00}
\definecolor{OrangeRed3}{rgb}{0.80,0.22,0.00}
\definecolor{OrangeRed4}{rgb}{0.55,0.15,0.00}
\definecolor{OrangeRed}{rgb}{1.00,0.27,0.00}
\definecolor{PaleGoldenrod}{rgb}{0.93,0.91,0.67}
\definecolor{PaleGreen1}{rgb}{0.60,1.00,0.60}
\definecolor{PaleGreen2}{rgb}{0.56,0.93,0.56}
\definecolor{PaleGreen3}{rgb}{0.49,0.80,0.49}
\definecolor{PaleGreen4}{rgb}{0.33,0.55,0.33}
\definecolor{PaleGreen}{rgb}{0.60,0.98,0.60}
\definecolor{PaleTurquoise1}{rgb}{0.73,1.00,1.00}
\definecolor{PaleTurquoise2}{rgb}{0.68,0.93,0.93}
\definecolor{PaleTurquoise3}{rgb}{0.59,0.80,0.80}
\definecolor{PaleTurquoise4}{rgb}{0.40,0.55,0.55}
\definecolor{PaleTurquoise}{rgb}{0.69,0.93,0.93}
\definecolor{PaleVioletRed1}{rgb}{1.00,0.51,0.67}
\definecolor{PaleVioletRed2}{rgb}{0.93,0.47,0.62}
\definecolor{PaleVioletRed3}{rgb}{0.80,0.41,0.54}
\definecolor{PaleVioletRed4}{rgb}{0.55,0.28,0.36}
\definecolor{PaleVioletRed}{rgb}{0.86,0.44,0.58}
\definecolor{PapayaWhip}{rgb}{1.00,0.94,0.84}
\definecolor{PeachPuff1}{rgb}{1.00,0.85,0.73}
\definecolor{PeachPuff2}{rgb}{0.93,0.80,0.68}
\definecolor{PeachPuff3}{rgb}{0.80,0.69,0.58}
\definecolor{PeachPuff4}{rgb}{0.55,0.47,0.40}
\definecolor{PeachPuff}{rgb}{1.00,0.85,0.73}
\definecolor{PowderBlue}{rgb}{0.69,0.88,0.90}
\definecolor{RosyBrown1}{rgb}{1.00,0.76,0.76}
\definecolor{RosyBrown2}{rgb}{0.93,0.71,0.71}
\definecolor{RosyBrown3}{rgb}{0.80,0.61,0.61}
\definecolor{RosyBrown4}{rgb}{0.55,0.41,0.41}
\definecolor{RosyBrown}{rgb}{0.74,0.56,0.56}
\definecolor{RoyalBlue1}{rgb}{0.28,0.46,1.00}
\definecolor{RoyalBlue2}{rgb}{0.26,0.43,0.93}
\definecolor{RoyalBlue3}{rgb}{0.23,0.37,0.80}
\definecolor{RoyalBlue4}{rgb}{0.15,0.25,0.55}
\definecolor{RoyalBlue}{rgb}{0.25,0.41,0.88}
\definecolor{SaddleBrown}{rgb}{0.55,0.27,0.07}
\definecolor{SandyBrown}{rgb}{0.96,0.64,0.38}
\definecolor{SeaGreen1}{rgb}{0.33,1.00,0.62}
\definecolor{SeaGreen2}{rgb}{0.31,0.93,0.58}
\definecolor{SeaGreen3}{rgb}{0.26,0.80,0.50}
\definecolor{SeaGreen4}{rgb}{0.18,0.55,0.34}
\definecolor{SeaGreen}{rgb}{0.18,0.55,0.34}
\definecolor{SkyBlue1}{rgb}{0.53,0.81,1.00}
\definecolor{SkyBlue2}{rgb}{0.49,0.75,0.93}
\definecolor{SkyBlue3}{rgb}{0.42,0.65,0.80}
\definecolor{SkyBlue4}{rgb}{0.29,0.44,0.55}
\definecolor{SkyBlue}{rgb}{0.53,0.81,0.92}
\definecolor{SlateBlue1}{rgb}{0.51,0.44,1.00}
\definecolor{SlateBlue2}{rgb}{0.48,0.40,0.93}
\definecolor{SlateBlue3}{rgb}{0.41,0.35,0.80}
\definecolor{SlateBlue4}{rgb}{0.28,0.24,0.55}
\definecolor{SlateBlue}{rgb}{0.42,0.35,0.80}
\definecolor{SlateGray1}{rgb}{0.78,0.89,1.00}
\definecolor{SlateGray2}{rgb}{0.73,0.83,0.93}
\definecolor{SlateGray3}{rgb}{0.62,0.71,0.80}
\definecolor{SlateGray4}{rgb}{0.42,0.48,0.55}
\definecolor{SlateGray}{rgb}{0.44,0.50,0.56}
\definecolor{SlateGrey}{rgb}{0.44,0.50,0.56}
\definecolor{SpringGreen1}{rgb}{0.00,1.00,0.50}
\definecolor{SpringGreen2}{rgb}{0.00,0.93,0.46}
\definecolor{SpringGreen3}{rgb}{0.00,0.80,0.40}
\definecolor{SpringGreen4}{rgb}{0.00,0.55,0.27}
\definecolor{SpringGreen}{rgb}{0.00,1.00,0.50}
\definecolor{SteelBlue1}{rgb}{0.39,0.72,1.00}
\definecolor{SteelBlue2}{rgb}{0.36,0.67,0.93}
\definecolor{SteelBlue3}{rgb}{0.31,0.58,0.80}
\definecolor{SteelBlue4}{rgb}{0.21,0.39,0.55}
\definecolor{SteelBlue}{rgb}{0.27,0.51,0.71}
\definecolor{VioletRed1}{rgb}{1.00,0.24,0.59}
\definecolor{VioletRed2}{rgb}{0.93,0.23,0.55}
\definecolor{VioletRed3}{rgb}{0.80,0.20,0.47}
\definecolor{VioletRed4}{rgb}{0.55,0.13,0.32}
\definecolor{VioletRed}{rgb}{0.82,0.13,0.56}
\definecolor{WhiteSmoke}{rgb}{0.96,0.96,0.96}
\definecolor{YellowGreen}{rgb}{0.60,0.80,0.20}
\definecolor{aliceblue}{rgb}{0.94,0.97,1.00}
\definecolor{antiquewhite}{rgb}{0.98,0.92,0.84}
\definecolor{aquamarine1}{rgb}{0.50,1.00,0.83}
\definecolor{aquamarine2}{rgb}{0.46,0.93,0.78}
\definecolor{aquamarine3}{rgb}{0.40,0.80,0.67}
\definecolor{aquamarine4}{rgb}{0.27,0.55,0.45}
\definecolor{aquamarine}{rgb}{0.50,1.00,0.83}
\definecolor{azure1}{rgb}{0.94,1.00,1.00}
\definecolor{azure2}{rgb}{0.88,0.93,0.93}
\definecolor{azure3}{rgb}{0.76,0.80,0.80}
\definecolor{azure4}{rgb}{0.51,0.55,0.55}
\definecolor{azure}{rgb}{0.94,1.00,1.00}
\definecolor{beige}{rgb}{0.96,0.96,0.86}
\definecolor{bisque1}{rgb}{1.00,0.89,0.77}
\definecolor{bisque2}{rgb}{0.93,0.84,0.72}
\definecolor{bisque3}{rgb}{0.80,0.72,0.62}
\definecolor{bisque4}{rgb}{0.55,0.49,0.42}
\definecolor{bisque}{rgb}{1.00,0.89,0.77}
\definecolor{black}{rgb}{0.00,0.00,0.00}
\definecolor{blanchedalmond}{rgb}{1.00,0.92,0.80}
\definecolor{blue1}{rgb}{0.00,0.00,1.00}
\definecolor{blue2}{rgb}{0.00,0.00,0.93}
\definecolor{blue3}{rgb}{0.00,0.00,0.80}
\definecolor{blue4}{rgb}{0.00,0.00,0.55}
\definecolor{blueviolet}{rgb}{0.54,0.17,0.89}
\definecolor{blue}{rgb}{0.00,0.00,1.00}
\definecolor{brown1}{rgb}{1.00,0.25,0.25}
\definecolor{brown2}{rgb}{0.93,0.23,0.23}
\definecolor{brown3}{rgb}{0.80,0.20,0.20}
\definecolor{brown4}{rgb}{0.55,0.14,0.14}
\definecolor{brown}{rgb}{0.65,0.16,0.16}
\definecolor{burlywood1}{rgb}{1.00,0.83,0.61}
\definecolor{burlywood2}{rgb}{0.93,0.77,0.57}
\definecolor{burlywood3}{rgb}{0.80,0.67,0.49}
\definecolor{burlywood4}{rgb}{0.55,0.45,0.33}
\definecolor{burlywood}{rgb}{0.87,0.72,0.53}
\definecolor{cadetblue}{rgb}{0.37,0.62,0.63}
\definecolor{chartreuse1}{rgb}{0.50,1.00,0.00}
\definecolor{chartreuse2}{rgb}{0.46,0.93,0.00}
\definecolor{chartreuse3}{rgb}{0.40,0.80,0.00}
\definecolor{chartreuse4}{rgb}{0.27,0.55,0.00}
\definecolor{chartreuse}{rgb}{0.50,1.00,0.00}
\definecolor{chocolate1}{rgb}{1.00,0.50,0.14}
\definecolor{chocolate2}{rgb}{0.93,0.46,0.13}
\definecolor{chocolate3}{rgb}{0.80,0.40,0.11}
\definecolor{chocolate4}{rgb}{0.55,0.27,0.07}
\definecolor{chocolate}{rgb}{0.82,0.41,0.12}
\definecolor{coral1}{rgb}{1.00,0.45,0.34}
\definecolor{coral2}{rgb}{0.93,0.42,0.31}
\definecolor{coral3}{rgb}{0.80,0.36,0.27}
\definecolor{coral4}{rgb}{0.55,0.24,0.18}
\definecolor{coral}{rgb}{1.00,0.50,0.31}
\definecolor{cornflowerblue}{rgb}{0.39,0.58,0.93}
\definecolor{cornsilk1}{rgb}{1.00,0.97,0.86}
\definecolor{cornsilk2}{rgb}{0.93,0.91,0.80}
\definecolor{cornsilk3}{rgb}{0.80,0.78,0.69}
\definecolor{cornsilk4}{rgb}{0.55,0.53,0.47}
\definecolor{cornsilk}{rgb}{1.00,0.97,0.86}
\definecolor{cyan1}{rgb}{0.00,1.00,1.00}
\definecolor{cyan2}{rgb}{0.00,0.93,0.93}
\definecolor{cyan3}{rgb}{0.00,0.80,0.80}
\definecolor{cyan4}{rgb}{0.00,0.55,0.55}
\definecolor{cyan}{rgb}{0.00,1.00,1.00}
\definecolor{darkblue}{rgb}{0.00,0.00,0.55}
\definecolor{darkcyan}{rgb}{0.00,0.55,0.55}
\definecolor{darkgoldenrod}{rgb}{0.72,0.53,0.04}
\definecolor{darkgray}{rgb}{0.66,0.66,0.66}
\definecolor{darkgreen}{rgb}{0.00,0.39,0.00}
\definecolor{darkgrey}{rgb}{0.66,0.66,0.66}
\definecolor{darkkhaki}{rgb}{0.74,0.72,0.42}
\definecolor{darkmagenta}{rgb}{0.55,0.00,0.55}
\definecolor{darkolive}{rgb}{0.33,0.42,0.18}
\definecolor{darkorange}{rgb}{1.00,0.55,0.00}
\definecolor{darkorchid}{rgb}{0.60,0.20,0.80}
\definecolor{darkred}{rgb}{0.55,0.00,0.00}
\definecolor{darksalmon}{rgb}{0.91,0.59,0.48}
\definecolor{darksea}{rgb}{0.56,0.74,0.56}
\definecolor{darkslate}{rgb}{0.18,0.31,0.31}
\definecolor{darkslate}{rgb}{0.18,0.31,0.31}
\definecolor{darkslate}{rgb}{0.28,0.24,0.55}
\definecolor{darkturquoise}{rgb}{0.00,0.81,0.82}
\definecolor{darkviolet}{rgb}{0.58,0.00,0.83}
\definecolor{deeppink}{rgb}{1.00,0.08,0.58}
\definecolor{deepsky}{rgb}{0.00,0.75,1.00}
\definecolor{dimgray}{rgb}{0.41,0.41,0.41}
\definecolor{dimgrey}{rgb}{0.41,0.41,0.41}
\definecolor{dodgerblue}{rgb}{0.12,0.56,1.00}
\definecolor{firebrick1}{rgb}{1.00,0.19,0.19}
\definecolor{firebrick2}{rgb}{0.93,0.17,0.17}
\definecolor{firebrick3}{rgb}{0.80,0.15,0.15}
\definecolor{firebrick4}{rgb}{0.55,0.10,0.10}
\definecolor{firebrick}{rgb}{0.70,0.13,0.13}
\definecolor{floralwhite}{rgb}{1.00,0.98,0.94}
\definecolor{forestgreen}{rgb}{0.13,0.55,0.13}
\definecolor{gainsboro}{rgb}{0.86,0.86,0.86}
\definecolor{ghostwhite}{rgb}{0.97,0.97,1.00}
\definecolor{gold1}{rgb}{1.00,0.84,0.00}
\definecolor{gold2}{rgb}{0.93,0.79,0.00}
\definecolor{gold3}{rgb}{0.80,0.68,0.00}
\definecolor{gold4}{rgb}{0.55,0.46,0.00}
\definecolor{goldenrod1}{rgb}{1.00,0.76,0.15}
\definecolor{goldenrod2}{rgb}{0.93,0.71,0.13}
\definecolor{goldenrod3}{rgb}{0.80,0.61,0.11}
\definecolor{goldenrod4}{rgb}{0.55,0.41,0.08}
\definecolor{goldenrod}{rgb}{0.85,0.65,0.13}
\definecolor{gold}{rgb}{1.00,0.84,0.00}
\definecolor{gray0}{rgb}{0.00,0.00,0.00}
\definecolor{gray100}{rgb}{1.00,1.00,1.00}
\definecolor{gray10}{rgb}{0.10,0.10,0.10}
\definecolor{gray11}{rgb}{0.11,0.11,0.11}
\definecolor{gray12}{rgb}{0.12,0.12,0.12}
\definecolor{gray13}{rgb}{0.13,0.13,0.13}
\definecolor{gray14}{rgb}{0.14,0.14,0.14}
\definecolor{gray15}{rgb}{0.15,0.15,0.15}
\definecolor{gray16}{rgb}{0.16,0.16,0.16}
\definecolor{gray17}{rgb}{0.17,0.17,0.17}
\definecolor{gray18}{rgb}{0.18,0.18,0.18}
\definecolor{gray19}{rgb}{0.19,0.19,0.19}
\definecolor{gray1}{rgb}{0.01,0.01,0.01}
\definecolor{gray20}{rgb}{0.20,0.20,0.20}
\definecolor{gray21}{rgb}{0.21,0.21,0.21}
\definecolor{gray22}{rgb}{0.22,0.22,0.22}
\definecolor{gray23}{rgb}{0.23,0.23,0.23}
\definecolor{gray24}{rgb}{0.24,0.24,0.24}
\definecolor{gray25}{rgb}{0.25,0.25,0.25}
\definecolor{gray26}{rgb}{0.26,0.26,0.26}
\definecolor{gray27}{rgb}{0.27,0.27,0.27}
\definecolor{gray28}{rgb}{0.28,0.28,0.28}
\definecolor{gray29}{rgb}{0.29,0.29,0.29}
\definecolor{gray2}{rgb}{0.02,0.02,0.02}
\definecolor{gray30}{rgb}{0.30,0.30,0.30}
\definecolor{gray31}{rgb}{0.31,0.31,0.31}
\definecolor{gray32}{rgb}{0.32,0.32,0.32}
\definecolor{gray33}{rgb}{0.33,0.33,0.33}
\definecolor{gray34}{rgb}{0.34,0.34,0.34}
\definecolor{gray35}{rgb}{0.35,0.35,0.35}
\definecolor{gray36}{rgb}{0.36,0.36,0.36}
\definecolor{gray37}{rgb}{0.37,0.37,0.37}
\definecolor{gray38}{rgb}{0.38,0.38,0.38}
\definecolor{gray39}{rgb}{0.39,0.39,0.39}
\definecolor{gray3}{rgb}{0.03,0.03,0.03}
\definecolor{gray40}{rgb}{0.40,0.40,0.40}
\definecolor{gray41}{rgb}{0.41,0.41,0.41}
\definecolor{gray42}{rgb}{0.42,0.42,0.42}
\definecolor{gray43}{rgb}{0.43,0.43,0.43}
\definecolor{gray44}{rgb}{0.44,0.44,0.44}
\definecolor{gray45}{rgb}{0.45,0.45,0.45}
\definecolor{gray46}{rgb}{0.46,0.46,0.46}
\definecolor{gray47}{rgb}{0.47,0.47,0.47}
\definecolor{gray48}{rgb}{0.48,0.48,0.48}
\definecolor{gray49}{rgb}{0.49,0.49,0.49}
\definecolor{gray4}{rgb}{0.04,0.04,0.04}
\definecolor{gray50}{rgb}{0.50,0.50,0.50}
\definecolor{gray51}{rgb}{0.51,0.51,0.51}
\definecolor{gray52}{rgb}{0.52,0.52,0.52}
\definecolor{gray53}{rgb}{0.53,0.53,0.53}
\definecolor{gray54}{rgb}{0.54,0.54,0.54}
\definecolor{gray55}{rgb}{0.55,0.55,0.55}
\definecolor{gray56}{rgb}{0.56,0.56,0.56}
\definecolor{gray57}{rgb}{0.57,0.57,0.57}
\definecolor{gray58}{rgb}{0.58,0.58,0.58}
\definecolor{gray59}{rgb}{0.59,0.59,0.59}
\definecolor{gray5}{rgb}{0.05,0.05,0.05}
\definecolor{gray60}{rgb}{0.60,0.60,0.60}
\definecolor{gray61}{rgb}{0.61,0.61,0.61}
\definecolor{gray62}{rgb}{0.62,0.62,0.62}
\definecolor{gray63}{rgb}{0.63,0.63,0.63}
\definecolor{gray64}{rgb}{0.64,0.64,0.64}
\definecolor{gray65}{rgb}{0.65,0.65,0.65}
\definecolor{gray66}{rgb}{0.66,0.66,0.66}
\definecolor{gray67}{rgb}{0.67,0.67,0.67}
\definecolor{gray68}{rgb}{0.68,0.68,0.68}
\definecolor{gray69}{rgb}{0.69,0.69,0.69}
\definecolor{gray6}{rgb}{0.06,0.06,0.06}
\definecolor{gray70}{rgb}{0.70,0.70,0.70}
\definecolor{gray71}{rgb}{0.71,0.71,0.71}
\definecolor{gray72}{rgb}{0.72,0.72,0.72}
\definecolor{gray73}{rgb}{0.73,0.73,0.73}
\definecolor{gray74}{rgb}{0.74,0.74,0.74}
\definecolor{gray75}{rgb}{0.75,0.75,0.75}
\definecolor{gray76}{rgb}{0.76,0.76,0.76}
\definecolor{gray77}{rgb}{0.77,0.77,0.77}
\definecolor{gray78}{rgb}{0.78,0.78,0.78}
\definecolor{gray79}{rgb}{0.79,0.79,0.79}
\definecolor{gray7}{rgb}{0.07,0.07,0.07}
\definecolor{gray80}{rgb}{0.80,0.80,0.80}
\definecolor{gray81}{rgb}{0.81,0.81,0.81}
\definecolor{gray82}{rgb}{0.82,0.82,0.82}
\definecolor{gray83}{rgb}{0.83,0.83,0.83}
\definecolor{gray84}{rgb}{0.84,0.84,0.84}
\definecolor{gray85}{rgb}{0.85,0.85,0.85}
\definecolor{gray86}{rgb}{0.86,0.86,0.86}
\definecolor{gray87}{rgb}{0.87,0.87,0.87}
\definecolor{gray88}{rgb}{0.88,0.88,0.88}
\definecolor{gray89}{rgb}{0.89,0.89,0.89}
\definecolor{gray8}{rgb}{0.08,0.08,0.08}
\definecolor{gray90}{rgb}{0.90,0.90,0.90}
\definecolor{gray91}{rgb}{0.91,0.91,0.91}
\definecolor{gray92}{rgb}{0.92,0.92,0.92}
\definecolor{gray93}{rgb}{0.93,0.93,0.93}
\definecolor{gray94}{rgb}{0.94,0.94,0.94}
\definecolor{gray95}{rgb}{0.95,0.95,0.95}
\definecolor{gray96}{rgb}{0.96,0.96,0.96}
\definecolor{gray97}{rgb}{0.97,0.97,0.97}
\definecolor{gray98}{rgb}{0.98,0.98,0.98}
\definecolor{gray99}{rgb}{0.99,0.99,0.99}
\definecolor{gray9}{rgb}{0.09,0.09,0.09}
\definecolor{gray}{rgb}{0.75,0.75,0.75}
\definecolor{green1}{rgb}{0.00,1.00,0.00}
\definecolor{green2}{rgb}{0.00,0.93,0.00}
\definecolor{green3}{rgb}{0.00,0.80,0.00}
\definecolor{green4}{rgb}{0.00,0.55,0.00}
\definecolor{greenyellow}{rgb}{0.68,1.00,0.18}
\definecolor{green}{rgb}{0.00,1.00,0.00}
\definecolor{grey0}{rgb}{0.00,0.00,0.00}
\definecolor{grey100}{rgb}{1.00,1.00,1.00}
\definecolor{grey10}{rgb}{0.10,0.10,0.10}
\definecolor{grey11}{rgb}{0.11,0.11,0.11}
\definecolor{grey12}{rgb}{0.12,0.12,0.12}
\definecolor{grey13}{rgb}{0.13,0.13,0.13}
\definecolor{grey14}{rgb}{0.14,0.14,0.14}
\definecolor{grey15}{rgb}{0.15,0.15,0.15}
\definecolor{grey16}{rgb}{0.16,0.16,0.16}
\definecolor{grey17}{rgb}{0.17,0.17,0.17}
\definecolor{grey18}{rgb}{0.18,0.18,0.18}
\definecolor{grey19}{rgb}{0.19,0.19,0.19}
\definecolor{grey1}{rgb}{0.01,0.01,0.01}
\definecolor{grey20}{rgb}{0.20,0.20,0.20}
\definecolor{grey21}{rgb}{0.21,0.21,0.21}
\definecolor{grey22}{rgb}{0.22,0.22,0.22}
\definecolor{grey23}{rgb}{0.23,0.23,0.23}
\definecolor{grey24}{rgb}{0.24,0.24,0.24}
\definecolor{grey25}{rgb}{0.25,0.25,0.25}
\definecolor{grey26}{rgb}{0.26,0.26,0.26}
\definecolor{grey27}{rgb}{0.27,0.27,0.27}
\definecolor{grey28}{rgb}{0.28,0.28,0.28}
\definecolor{grey29}{rgb}{0.29,0.29,0.29}
\definecolor{grey2}{rgb}{0.02,0.02,0.02}
\definecolor{grey30}{rgb}{0.30,0.30,0.30}
\definecolor{grey31}{rgb}{0.31,0.31,0.31}
\definecolor{grey32}{rgb}{0.32,0.32,0.32}
\definecolor{grey33}{rgb}{0.33,0.33,0.33}
\definecolor{grey34}{rgb}{0.34,0.34,0.34}
\definecolor{grey35}{rgb}{0.35,0.35,0.35}
\definecolor{grey36}{rgb}{0.36,0.36,0.36}
\definecolor{grey37}{rgb}{0.37,0.37,0.37}
\definecolor{grey38}{rgb}{0.38,0.38,0.38}
\definecolor{grey39}{rgb}{0.39,0.39,0.39}
\definecolor{grey3}{rgb}{0.03,0.03,0.03}
\definecolor{grey40}{rgb}{0.40,0.40,0.40}
\definecolor{grey41}{rgb}{0.41,0.41,0.41}
\definecolor{grey42}{rgb}{0.42,0.42,0.42}
\definecolor{grey43}{rgb}{0.43,0.43,0.43}
\definecolor{grey44}{rgb}{0.44,0.44,0.44}
\definecolor{grey45}{rgb}{0.45,0.45,0.45}
\definecolor{grey46}{rgb}{0.46,0.46,0.46}
\definecolor{grey47}{rgb}{0.47,0.47,0.47}
\definecolor{grey48}{rgb}{0.48,0.48,0.48}
\definecolor{grey49}{rgb}{0.49,0.49,0.49}
\definecolor{grey4}{rgb}{0.04,0.04,0.04}
\definecolor{grey50}{rgb}{0.50,0.50,0.50}
\definecolor{grey51}{rgb}{0.51,0.51,0.51}
\definecolor{grey52}{rgb}{0.52,0.52,0.52}
\definecolor{grey53}{rgb}{0.53,0.53,0.53}
\definecolor{grey54}{rgb}{0.54,0.54,0.54}
\definecolor{grey55}{rgb}{0.55,0.55,0.55}
\definecolor{grey56}{rgb}{0.56,0.56,0.56}
\definecolor{grey57}{rgb}{0.57,0.57,0.57}
\definecolor{grey58}{rgb}{0.58,0.58,0.58}
\definecolor{grey59}{rgb}{0.59,0.59,0.59}
\definecolor{grey5}{rgb}{0.05,0.05,0.05}
\definecolor{grey60}{rgb}{0.60,0.60,0.60}
\definecolor{grey61}{rgb}{0.61,0.61,0.61}
\definecolor{grey62}{rgb}{0.62,0.62,0.62}
\definecolor{grey63}{rgb}{0.63,0.63,0.63}
\definecolor{grey64}{rgb}{0.64,0.64,0.64}
\definecolor{grey65}{rgb}{0.65,0.65,0.65}
\definecolor{grey66}{rgb}{0.66,0.66,0.66}
\definecolor{grey67}{rgb}{0.67,0.67,0.67}
\definecolor{grey68}{rgb}{0.68,0.68,0.68}
\definecolor{grey69}{rgb}{0.69,0.69,0.69}
\definecolor{grey6}{rgb}{0.06,0.06,0.06}
\definecolor{grey70}{rgb}{0.70,0.70,0.70}
\definecolor{grey71}{rgb}{0.71,0.71,0.71}
\definecolor{grey72}{rgb}{0.72,0.72,0.72}
\definecolor{grey73}{rgb}{0.73,0.73,0.73}
\definecolor{grey74}{rgb}{0.74,0.74,0.74}
\definecolor{grey75}{rgb}{0.75,0.75,0.75}
\definecolor{grey76}{rgb}{0.76,0.76,0.76}
\definecolor{grey77}{rgb}{0.77,0.77,0.77}
\definecolor{grey78}{rgb}{0.78,0.78,0.78}
\definecolor{grey79}{rgb}{0.79,0.79,0.79}
\definecolor{grey7}{rgb}{0.07,0.07,0.07}
\definecolor{grey80}{rgb}{0.80,0.80,0.80}
\definecolor{grey81}{rgb}{0.81,0.81,0.81}
\definecolor{grey82}{rgb}{0.82,0.82,0.82}
\definecolor{grey83}{rgb}{0.83,0.83,0.83}
\definecolor{grey84}{rgb}{0.84,0.84,0.84}
\definecolor{grey85}{rgb}{0.85,0.85,0.85}
\definecolor{grey86}{rgb}{0.86,0.86,0.86}
\definecolor{grey87}{rgb}{0.87,0.87,0.87}
\definecolor{grey88}{rgb}{0.88,0.88,0.88}
\definecolor{grey89}{rgb}{0.89,0.89,0.89}
\definecolor{grey8}{rgb}{0.08,0.08,0.08}
\definecolor{grey90}{rgb}{0.90,0.90,0.90}
\definecolor{grey91}{rgb}{0.91,0.91,0.91}
\definecolor{grey92}{rgb}{0.92,0.92,0.92}
\definecolor{grey93}{rgb}{0.93,0.93,0.93}
\definecolor{grey94}{rgb}{0.94,0.94,0.94}
\definecolor{grey95}{rgb}{0.95,0.95,0.95}
\definecolor{grey96}{rgb}{0.96,0.96,0.96}
\definecolor{grey97}{rgb}{0.97,0.97,0.97}
\definecolor{grey98}{rgb}{0.98,0.98,0.98}
\definecolor{grey99}{rgb}{0.99,0.99,0.99}
\definecolor{grey9}{rgb}{0.09,0.09,0.09}
\definecolor{grey}{rgb}{0.75,0.75,0.75}
\definecolor{honeydew1}{rgb}{0.94,1.00,0.94}
\definecolor{honeydew2}{rgb}{0.88,0.93,0.88}
\definecolor{honeydew3}{rgb}{0.76,0.80,0.76}
\definecolor{honeydew4}{rgb}{0.51,0.55,0.51}
\definecolor{honeydew}{rgb}{0.94,1.00,0.94}
\definecolor{hotpink}{rgb}{1.00,0.41,0.71}
\definecolor{indianred}{rgb}{0.80,0.36,0.36}
\definecolor{ivory1}{rgb}{1.00,1.00,0.94}
\definecolor{ivory2}{rgb}{0.93,0.93,0.88}
\definecolor{ivory3}{rgb}{0.80,0.80,0.76}
\definecolor{ivory4}{rgb}{0.55,0.55,0.51}
\definecolor{ivory}{rgb}{1.00,1.00,0.94}
\definecolor{khaki1}{rgb}{1.00,0.96,0.56}
\definecolor{khaki2}{rgb}{0.93,0.90,0.52}
\definecolor{khaki3}{rgb}{0.80,0.78,0.45}
\definecolor{khaki4}{rgb}{0.55,0.53,0.31}
\definecolor{khaki}{rgb}{0.94,0.90,0.55}
\definecolor{lavenderblush}{rgb}{1.00,0.94,0.96}
\definecolor{lavender}{rgb}{0.90,0.90,0.98}
\definecolor{lawngreen}{rgb}{0.49,0.99,0.00}
\definecolor{lemonchiffon}{rgb}{1.00,0.98,0.80}
\definecolor{lightblue}{rgb}{0.68,0.85,0.90}
\definecolor{lightcoral}{rgb}{0.94,0.50,0.50}
\definecolor{lightcyan}{rgb}{0.88,1.00,1.00}
\definecolor{lightgoldenrod}{rgb}{0.93,0.87,0.51}
\definecolor{lightgoldenrod}{rgb}{0.98,0.98,0.82}
\definecolor{lightgray}{rgb}{0.83,0.83,0.83}
\definecolor{lightgreen}{rgb}{0.56,0.93,0.56}
\definecolor{lightgrey}{rgb}{0.83,0.83,0.83}
\definecolor{lightpink}{rgb}{1.00,0.71,0.76}
\definecolor{lightsalmon}{rgb}{1.00,0.63,0.48}
\definecolor{lightsea}{rgb}{0.13,0.70,0.67}
\definecolor{lightsky}{rgb}{0.53,0.81,0.98}
\definecolor{lightslate}{rgb}{0.47,0.53,0.60}
\definecolor{lightslate}{rgb}{0.47,0.53,0.60}
\definecolor{lightslate}{rgb}{0.52,0.44,1.00}
\definecolor{lightsteel}{rgb}{0.69,0.77,0.87}
\definecolor{lightyellow}{rgb}{1.00,1.00,0.88}
\definecolor{limegreen}{rgb}{0.20,0.80,0.20}
\definecolor{linen}{rgb}{0.98,0.94,0.90}
\definecolor{magenta1}{rgb}{1.00,0.00,1.00}
\definecolor{magenta2}{rgb}{0.93,0.00,0.93}
\definecolor{magenta3}{rgb}{0.80,0.00,0.80}
\definecolor{magenta4}{rgb}{0.55,0.00,0.55}
\definecolor{magenta}{rgb}{1.00,0.00,1.00}
\definecolor{maroon1}{rgb}{1.00,0.20,0.70}
\definecolor{maroon2}{rgb}{0.93,0.19,0.65}
\definecolor{maroon3}{rgb}{0.80,0.16,0.56}
\definecolor{maroon4}{rgb}{0.55,0.11,0.38}
\definecolor{maroon}{rgb}{0.69,0.19,0.38}
\definecolor{mediumaquamarine}{rgb}{0.40,0.80,0.67}
\definecolor{mediumblue}{rgb}{0.00,0.00,0.80}
\definecolor{mediumorchid}{rgb}{0.73,0.33,0.83}
\definecolor{mediumpurple}{rgb}{0.58,0.44,0.86}
\definecolor{mediumsea}{rgb}{0.24,0.70,0.44}
\definecolor{mediumslate}{rgb}{0.48,0.41,0.93}
\definecolor{mediumspring}{rgb}{0.00,0.98,0.60}
\definecolor{mediumturquoise}{rgb}{0.28,0.82,0.80}
\definecolor{mediumviolet}{rgb}{0.78,0.08,0.52}
\definecolor{midnightblue}{rgb}{0.10,0.10,0.44}
\definecolor{mintcream}{rgb}{0.96,1.00,0.98}
\definecolor{mistyrose}{rgb}{1.00,0.89,0.88}
\definecolor{moccasin}{rgb}{1.00,0.89,0.71}
\definecolor{navajowhite}{rgb}{1.00,0.87,0.68}
\definecolor{navyblue}{rgb}{0.00,0.00,0.50}
\definecolor{navy}{rgb}{0.00,0.00,0.50}
\definecolor{oldlace}{rgb}{0.99,0.96,0.90}
\definecolor{olivedrab}{rgb}{0.42,0.56,0.14}
\definecolor{orange1}{rgb}{1.00,0.65,0.00}
\definecolor{orange2}{rgb}{0.93,0.60,0.00}
\definecolor{orange3}{rgb}{0.80,0.52,0.00}
\definecolor{orange4}{rgb}{0.55,0.35,0.00}
\definecolor{orangered}{rgb}{1.00,0.27,0.00}
\definecolor{orange}{rgb}{1.00,0.65,0.00}
\definecolor{orchid1}{rgb}{1.00,0.51,0.98}
\definecolor{orchid2}{rgb}{0.93,0.48,0.91}
\definecolor{orchid3}{rgb}{0.80,0.41,0.79}
\definecolor{orchid4}{rgb}{0.55,0.28,0.54}
\definecolor{orchid}{rgb}{0.85,0.44,0.84}
\definecolor{palegoldenrod}{rgb}{0.93,0.91,0.67}
\definecolor{palegreen}{rgb}{0.60,0.98,0.60}
\definecolor{paleturquoise}{rgb}{0.69,0.93,0.93}
\definecolor{paleviolet}{rgb}{0.86,0.44,0.58}
\definecolor{papayawhip}{rgb}{1.00,0.94,0.84}
\definecolor{peachpuff}{rgb}{1.00,0.85,0.73}
\definecolor{peru}{rgb}{0.80,0.52,0.25}
\definecolor{pink1}{rgb}{1.00,0.71,0.77}
\definecolor{pink2}{rgb}{0.93,0.66,0.72}
\definecolor{pink3}{rgb}{0.80,0.57,0.62}
\definecolor{pink4}{rgb}{0.55,0.39,0.42}
\definecolor{pink}{rgb}{1.00,0.75,0.80}
\definecolor{plum1}{rgb}{1.00,0.73,1.00}
\definecolor{plum2}{rgb}{0.93,0.68,0.93}
\definecolor{plum3}{rgb}{0.80,0.59,0.80}
\definecolor{plum4}{rgb}{0.55,0.40,0.55}
\definecolor{plum}{rgb}{0.87,0.63,0.87}
\definecolor{powderblue}{rgb}{0.69,0.88,0.90}
\definecolor{purple1}{rgb}{0.61,0.19,1.00}
\definecolor{purple2}{rgb}{0.57,0.17,0.93}
\definecolor{purple3}{rgb}{0.49,0.15,0.80}
\definecolor{purple4}{rgb}{0.33,0.10,0.55}
\definecolor{purple}{rgb}{0.63,0.13,0.94}
\definecolor{red1}{rgb}{1.00,0.00,0.00}
\definecolor{red2}{rgb}{0.93,0.00,0.00}
\definecolor{red3}{rgb}{0.80,0.00,0.00}
\definecolor{red4}{rgb}{0.55,0.00,0.00}
\definecolor{red}{rgb}{1.00,0.00,0.00}
\definecolor{rosybrown}{rgb}{0.74,0.56,0.56}
\definecolor{royalblue}{rgb}{0.25,0.41,0.88}
\definecolor{saddlebrown}{rgb}{0.55,0.27,0.07}
\definecolor{salmon1}{rgb}{1.00,0.55,0.41}
\definecolor{salmon2}{rgb}{0.93,0.51,0.38}
\definecolor{salmon3}{rgb}{0.80,0.44,0.33}
\definecolor{salmon4}{rgb}{0.55,0.30,0.22}
\definecolor{salmon}{rgb}{0.98,0.50,0.45}
\definecolor{sandybrown}{rgb}{0.96,0.64,0.38}
\definecolor{seagreen}{rgb}{0.18,0.55,0.34}
\definecolor{seashell1}{rgb}{1.00,0.96,0.93}
\definecolor{seashell2}{rgb}{0.93,0.90,0.87}
\definecolor{seashell3}{rgb}{0.80,0.77,0.75}
\definecolor{seashell4}{rgb}{0.55,0.53,0.51}
\definecolor{seashell}{rgb}{1.00,0.96,0.93}
\definecolor{sienna1}{rgb}{1.00,0.51,0.28}
\definecolor{sienna2}{rgb}{0.93,0.47,0.26}
\definecolor{sienna3}{rgb}{0.80,0.41,0.22}
\definecolor{sienna4}{rgb}{0.55,0.28,0.15}
\definecolor{sienna}{rgb}{0.63,0.32,0.18}
\definecolor{skyblue}{rgb}{0.53,0.81,0.92}
\definecolor{slateblue}{rgb}{0.42,0.35,0.80}
\definecolor{slategray}{rgb}{0.44,0.50,0.56}
\definecolor{slategrey}{rgb}{0.44,0.50,0.56}
\definecolor{snow1}{rgb}{1.00,0.98,0.98}
\definecolor{snow2}{rgb}{0.93,0.91,0.91}
\definecolor{snow3}{rgb}{0.80,0.79,0.79}
\definecolor{snow4}{rgb}{0.55,0.54,0.54}
\definecolor{snow}{rgb}{1.00,0.98,0.98}
\definecolor{springgreen}{rgb}{0.00,1.00,0.50}
\definecolor{steelblue}{rgb}{0.27,0.51,0.71}
\definecolor{tan1}{rgb}{1.00,0.65,0.31}
\definecolor{tan2}{rgb}{0.93,0.60,0.29}
\definecolor{tan3}{rgb}{0.80,0.52,0.25}
\definecolor{tan4}{rgb}{0.55,0.35,0.17}
\definecolor{tan}{rgb}{0.82,0.71,0.55}
\definecolor{thistle1}{rgb}{1.00,0.88,1.00}
\definecolor{thistle2}{rgb}{0.93,0.82,0.93}
\definecolor{thistle3}{rgb}{0.80,0.71,0.80}
\definecolor{thistle4}{rgb}{0.55,0.48,0.55}
\definecolor{thistle}{rgb}{0.85,0.75,0.85}
\definecolor{tomato1}{rgb}{1.00,0.39,0.28}
\definecolor{tomato2}{rgb}{0.93,0.36,0.26}
\definecolor{tomato3}{rgb}{0.80,0.31,0.22}
\definecolor{tomato4}{rgb}{0.55,0.21,0.15}
\definecolor{tomato}{rgb}{1.00,0.39,0.28}
\definecolor{turquoise1}{rgb}{0.00,0.96,1.00}
\definecolor{turquoise2}{rgb}{0.00,0.90,0.93}
\definecolor{turquoise3}{rgb}{0.00,0.77,0.80}
\definecolor{turquoise4}{rgb}{0.00,0.53,0.55}
\definecolor{turquoise}{rgb}{0.25,0.88,0.82}
\definecolor{violetred}{rgb}{0.82,0.13,0.56}
\definecolor{violet}{rgb}{0.93,0.51,0.93}
\definecolor{wheat1}{rgb}{1.00,0.91,0.73}
\definecolor{wheat2}{rgb}{0.93,0.85,0.68}
\definecolor{wheat3}{rgb}{0.80,0.73,0.59}
\definecolor{wheat4}{rgb}{0.55,0.49,0.40}
\definecolor{wheat}{rgb}{0.96,0.87,0.70}
\definecolor{whitesmoke}{rgb}{0.96,0.96,0.96}
\definecolor{white}{rgb}{1.00,1.00,1.00}
\definecolor{yellow1}{rgb}{1.00,1.00,0.00}
\definecolor{yellow2}{rgb}{0.93,0.93,0.00}
\definecolor{yellow3}{rgb}{0.80,0.80,0.00}
\definecolor{yellow4}{rgb}{0.55,0.55,0.00}
\definecolor{yellowgreen}{rgb}{0.60,0.80,0.20}
\definecolor{yellow}{rgb}{1.00,1.00,0.00}

\usepackage{xcolor,colortbl}
\definecolor{red}{rgb}{1,0.3,0.2}
\newcolumntype{m}{>{\columncolor{red}}c}
\newcolumntype{b}{>{\columncolor{red}}c}

\newcommand\swift{{\it Swift}}

\newcommand\rosat{{\it ROSAT}}
\newcommand\rxte{{\it RXTE}}
\newcommand\iue{{\it International Ultraviolet Explorer}}
\newcommand\xmm{{\it XMM-Newton}}

\newcommand\ks{{\rm~ks}}

\newcommand\mpc{{\rm~Mpc}}

\newcommand\kev{{\rm~keV}}
\newcommand\ev{{\rm~eV}}
\newcommand\kms{\ifmmode {\rm~km\ s}$^{-1}$ \else ~km s$^{-1}$\fi}
\newcommand\Hunit{\ifmmode {\rm~km\ s}$^{-1}$\ {\rm Mpc}$^{-1}$
        \else ~km s$^{-1}$ Mpc$^{-1}$\fi}
\newcommand\ctssec{\ifmmode {\rm~count\ s}$^{-1}$ \else ~count s$^{-1}$\fi}
\newcommand\ergsec{\ifmmode {\rm~erg\ s}$^{-1}$ \else
        ~erg s$^{-1}$\fi}
\newcommand\funit{\ifmmode {\rm~erg\ s}$^{-1}$\;{\rm cm}$^{-2}$ \else
        ~ergs s$^{-1}$ cm$^{-2}$\fi}
\newcommand\phflux{\ifmmode {\rm~photon\ s}$^{-1}$\;{\rm cm}$^{-i2}$
        \else   ~photon s$^{-1}$ cm$^{-2}$\fi}
\newcommand\efluxA{\ifmmode {\rm~erg\ s}$^{-1}$\;{\rm cm}$^{-2}$\;{\rm
        \AA}$^{-1}$ \else ~erg s$^{-1}$ cm$^{-2}$ \AA$^{-1}$\fi}
\newcommand\efluxHz{\ifmmode {\rm~erg\ s}$^{-1}$\;{\rm cm}$^{-2}$\;{\rm
        Hz}$^{-1}$ \else ~erg s$^{-1}$ cm$^{-2}$ Hz$^{-1}$\fi}
\newcommand\cc{\ifmmode {\rm~cm}$^{-3}$ \else cm$^{-3}$\fi}
\newcommand\FWHM{\ifmmode {\rm~FWHM} \else ${\rm~FWHM}$\fi}
\newcommand\Msun{\ifmmode M_{\odot} \else $M_{\odot}$\fi}
\newcommand\Lsun{\ifmmode L_{\odot} \else $L_{\odot}$\fi}

\newcommand\hbeta{\ifmmode {\rm H}\beta \else H$\beta$\fi}
\newcommand\Kalpha{\ifmmode {\rm K}\alpha \else K$\alpha$\fi}
\newcommand\nh{\ifmmode N_{\rm H} \else N$_{\rm H}$\fi}

\makeatletter

\newcommand{\Rmnum}[1]{\expandafter\@slowromancap\romannumeral #1@}
\makeatother

\title [Spectral variability of II~Zw~177]{X-ray/UV variability and the origin of soft X-ray excess emission from II~Zw~177}
\author[Pal et al.]{Main Pal$^{1}$\thanks{\ mainpal@iucaa.in}, Gulab~C.~Dewangan$^{1}$\footnotemark[1],  Ranjeev~Misra$^{1}$\footnotemark[1],
Pramod~K.~Pawar$^{2}$\\
$^{1}$Inter University Centre for Astronomy and Astrophysics, Pune 411 007, India.\\
$^{2}$Swami Ramanand Teerth Marathwada University, Nanded, India.}

\begin{document}
\pagerange{\pageref{firstpage}--\pageref{lastpage}} \pubyear{2012}
\date{\today}

\maketitle
\begin{abstract}
  We study X-ray and UV emission from the narrow-line Seyfert 1 galaxy
  II~Zw~177 using a $137\ks$ long and another $13\ks$ short \xmm{}
  observation performed in 2012 and 2001, respectively.  Both 
  observations show soft X-ray excess emission contributing
  $76.9\pm4.9\%$ in 2012 and $58.8\pm10.2\%$ in 2001 in the
  $0.3-2\kev$ band. We find that both blurred reflection from an
  ionized disc and Comptonized disc emission describe the
  observed soft excess well. Time-resolved spectroscopy
  on scales of $\sim20\ks$ reveals strong correlation between the soft
  excess and the powerlaw components. The fractional variability amplitude $F_{var}$ derived from EPIC-pn lightcurves at different energy bands is nearly constant ($F_{var} \sim20\%$). This is in contrast to other
  AGNs where the lack of short term variation in soft X-ray excess emission
  has been attributed to intense light bending in the framework of the
  ``lamppost''  model. Thus, the variations in powerlaw emission are most
  likely intrinsic to corona rather than just due to the changes of height
  of compact corona. The variable UV emission ($F_{var} \sim 1\%$) is
  uncorrelated to any of the X-ray components on short timescales suggesting
  that the UV emission is not dominated by the reprocessed emission. The
  gradual observed decline in the UV emission in 2012 may be related to
  the secular decline due to the changes in the accretion
   rate. In this case, the short term X-ray variability is
not due to the changes in the seed photons but intrinsic to the hot corona.
\end{abstract}

\section{Introduction}

\label{firstpage}

Most active galactic nuclei (AGN) show complex broad-band
spectra. The optical/UV to X-ray spectrum consist of primarily four
components -- the big blue bump (BBB), the X-ray power law emission in
the $\sim0.1-100\kev$ band, Fe-K line near 6\kev~with reflection hump
at higher energies and the soft X-ray excess emission below 2\kev.
The BBB is likely the thermal emission from a standard accretion disc
\citep{1973A&A....24..337S, 1978Natur.272..706S}.  According to this
model the outer accretion disc with low temperature emits within the optical band
and the inner disc with the highest temperature emits in the extreme UV.
 In rare cases of low mass AGN with high accretion rates, the disc
can emit in the soft X-ray band (e.g., \citealt{2013MNRAS.434.1955D, 2015MNRAS.446..759C}).
The power law continuum emission is believed to originate through the Compton
up scattering of the soft photons from the accretion disc in the hot electron plasma
\citep{1980A&A....86..121S, 1991ApJ...380L..51H, 2003PhR...377..389R}.
The Fe-K$\alpha$ line near 6\kev~and the reflection hump in the
$20-40$\kev~band are considered together as the X-ray reflection. The
iron-K$\alpha$ line and the reflection hump are caused by the
photoelectric absorption, followed by fluorescence, and the Compton
scattering of the coronal emission in an
optically thick medium such as accretion discs
\citep{1988MNRAS.233..475G, 1988ApJ...335...57L, 1991MNRAS.249..352G,
  1995Natur.375..659T}.  Since the discovery of the soft X-ray excess
emission \citep{1985ApJ...297..633S, 1985MNRAS.217..105A}, the
origin of this component has been widely debated with high energy
tail of thermal emission from the accretion disc discussed as an early possibility
\citep{1985MNRAS.217..105A,1998MNRAS.301..179M, 1999ApJS..125..317L}.
However, the expected thermal emission from accretion disc of AGN is too cool to
contribute significantly in the soft X-ray band, except for a subclass of AGN known as 
narrow line Seyfert galaxies (NLS1s) which have low black hole mass and high
accretion rate (e.g., \citealt{1996A&A...305...53B}). 

Nonetheless, in NLS1, the predicted shape of Wein tail of
inner disc should drop faster than the observed shape of soft excess with
energy \citep{1987ApJ...314..699B, 1997ApJ...477...93L}. The extent over the
Wein tail of the disc could be explained with Compton upscattering of disc photons
through optically thick cool plasma \citep{1987ApJ...321..305C}. 
This optically thick cool plasma may reside between the interior of disc and the X-ray source
either separated vertically \citep{2001ApJ...557..408J} or radially
\citep{1998MNRAS.301..179M} or indistinct \citep{2012MNRAS.420.1848D}. However,
the temperature associated with this Comptonization region remains invariant
$kT_{e}\sim0.1-0.2$\kev~\citep{2003A&A...412..317C,2004MNRAS.349L...7G,
2009MNRAS.394..443M}. This constancy of temperature over a large range in AGN
masses and luminosities led to the two models for soft excess -- blurred
reflection from relativistic photoionized accretion disc
\citep{2002MNRAS.331L..35F,2006MNRAS.365.1067C, 2013MNRAS.428.2901W} and smeared ionized absorption
\citep{2004MNRAS.349L...7G,2006MNRAS.371L..16G, 2006MNRAS.371...81S,
2007MNRAS.381.1426M} that utilized the atomic processes. The velocity smeared
absorption explanation for soft excess was ruled out due to the unrealistic
requirement of very high velocity ($\sim0.5 c$) of gas clouds to smooth the
curvature of this excess \citep{2008MNRAS.386L...1S, 2009ApJ...694....1S}.
However, the origin of soft X-ray excess is still not clearly understood since
these models are often spectroscopically degenerate in the sense that they fit
the X-ray spectra equally well (e.g., \citealt{2004MNRAS.349L...7G,
2005AIPC..774..317S, 2007ApJ...671.1284D}). The variability study based on
hard/soft time lags in different energy bands and correlation
between spectral components can help to remove this ambiguity between
different physical processes.

 After the discovery of hard X-ray lag relative to soft X-rays
	in binary system (e.g.,
	\citealt{1988Natur.336..450M,1999ApJ...510..874N}), similar lags were also
	observed in AGN \citep{2001ApJ...554L.133P,2004MNRAS.348..783M}. The
	origin of hard lag is not clearly known. One possible explanation
	is provided by the propagation fluctuation model in which fluctuations
	associated with accretion flow propagate inward in an accretion disc and
	thus resulting in the emission of the soft photons from relatively outer
	regions earlier than the hard photons from the innermost regions \citep{1997MNRAS.292..679L,2001MNRAS.327..799K, 2006MNRAS.367..801A}. 
	Recently, for example, \swift{} monitoring of the radio-loud NLS1 galaxy PKS~0558--504 for
$\sim 1.5{\rm~year}$ has revealed that optical leads UV and UV leads soft X-rays
on short timescales of about a week \citep{2013MNRAS.433.1709G} possibly
favouring the propagation model. A new type of lag has emerged from recent
studies where soft photons lag to the hard photons. This is termed as the
reverberation lag which is used to constrain the X-ray emitting region in AGN.
\citet{2009Natur.459..540F} discovered the reverberation lag $\sim$ 30s for the
first time in a NLS1 galaxy 1H0707--495. Since then such lags have been observed
in dozen of Seyfert galaxies \citep{2010MNRAS.401.2419Z, 2011MNRAS.417L..98D,
2011MNRAS.416L..94E,
2011MNRAS.412...59Z,2011MNRAS.418.2642Z,2013ApJ...764L...9C,
2013MNRAS.429.2917F, 2013MNRAS.428.2795K, 2013MNRAS.431.2441D}. The most of
above cases reveal reverberation lag $\sim100$s supporting the compact nature of
X-ray emitting region within few gravitational radii of a supermassive black
hole (SMBH).  In case of strong illumination such as that implied by observation of
strong blurred reflection, UV/optical emission from AGN may be dominated by the
reprocessed emission and the variations in the optical/UV  
band emission lag behind the X-rays (e.g., \citealt{2014MNRAS.444.1469M}). About a five year
long compaign of Seyfert  1 galaxy Mrk~79 using six ground based
observatories for optical and \rxte{} for X-ray observations,
\citet{2009MNRAS.394..427B} have shown zero lag between optical and X-rays on
timescale of about a day. Their study of correlated X-ray and optical emission
suggests X-ray reprocessing on short timescale of days and the changes in the
optical emission on long timescale of $\sim$ few years can be attributed to the
variations in the accretion rate.  

The presence or absence of
correlated variability between the different X-ray spectral components
can also be used to constrain the models. For example, one would
naively expect that the soft excess should be correlated with
the power-law emission for the reflection model. However, it has
been argued that strong gravitational light bending within the
framework of a ``lamppost'' model where a compact  X-ray producing region 
located at a height illuminates the disc, may lead to absence of such correlations
\citep{2004MNRAS.349.1435M}. It is also difficult to eliminate the
other models for the soft excess based on correlation studies of X-ray components
alone. Multiwavelength variability properties and the relationship between the soft X-ray
excess  and the optical/UV emission from the accretion disc may
provide better constraints on the models for the soft X-ray excess.

Multiwavelength studies of AGN have shown possible relationship between the soft X-ray excess and the optical/UV big blue bump
(e.g.,~\citealt{1992MNRAS.256..589P,1996A&A...305...53B, 1994ApJS...95....1E, 1996ApJ...470..364E,
2006MNRAS.366..953B, 2007MNRAS.381.1426M}). \citet{1993A&A...274..105W} found a strong correlation
between the soft X-ray slope measured with \rosat{} and the strength of UV emission observed with
\iue{} for 58 Seyfert 1 AGN. \citet{1998A&A...330...25G} found pronounced BBB emission in a sample
of 76 bright soft X-ray selected AGN and the optical spectra of these AGN were bluer when the soft
X-ray spectra were steeper. They interpreted these results as the BBB emission arising from
Comptonized accretion disc. Based on a multiwavelength campaign of Mrk~509 by
\swift{} and \xmm{} over a period of 100 days, \citet{2011A&A...534A..39M} have
found a strong correlation between the soft X-ray excess and the optical/UV thermal emission from the
accretion disc but no correlation between the X-ray powerlaw component
and either the soft X-ray excess or optical/UV emission. These results
are expected if the soft X-ray excess is produced in the thermal
Comptonization of optical/UV disc photons by a warm ($kT_e \sim
0.2\kev$) and optically thick ($\tau \sim 17$) corona surrounding
the inner disc \citep{2011A&A...534A..39M}. Thus, multiwavelength optical/UV/X-ray variability study of AGN is an
important tool to investigate the relationship between different
emission components and to shed light on the central engine and the
origin of soft X-ray excess. In this paper we present detailed
broadband X-ray spectral and UV/X-ray variability study of II~Zw~177
-- a highly variable AGN.

II~Zw~177 is a bright X-ray source and is listed in the \rosat{} X-ray
bright source catalogue \citep{1999A&A...349..389V}. Its luminosity
was measured $\sim10^{43}~ergs~s^{-1}$ in $0.5-2~\kev$~band with
\xmm{} \citep{2009A&A...495..421B}. II~Zw~177 is a NLS1 galaxy located
at a redshift $z=0.081$ \citep{1998ApJS..117..319A,
  2006A&A...455..773V, 2006MNRAS.368..479G, 2006ApJS..166..128Z,
  2011ApJ...727...31A}.  \citet{2006MNRAS.368..479G} studied this AGN
using 2001 June \xmm{} observation and grouped this AGN in the
``General NLS1 Sample'' as opposed to the ``Complex NLS1 Sample'' and
also reported a broad feature near $5.8\kev$.  We have performed a
long \xmm{} observation which we analyze here. We also use the earlier
observation and study X-ray spectral and UV/X-ray variability.  We
organize this paper as follows. We describe the observations and data
reduction in Section 2. In Section 3, we detail the spectral modeling
with complex models. In Section 4, we present the long term X-ray spectral variability.
We report the UV to X-ray spectral variability in Section 5.
We discuss and summarize our results in Section 6. We use the luminosity distance
($d_{L}=364\mpc$) using cosmological parameters $H_{0} = 71~{\rm
  km~s^{-1}~Mpc^{-1}}$, $\Omega_m = 0.27$ and $\Omega_{\Lambda} =
0.73$.

\begin{table*}
  \noindent \begin{centering}
    \caption{Observing log} \label{obs_log}
    \begin{tabular*}{12.5 cm}{lccccccc}
      \hline
      \hline 
      Observatory & Observation-ID&Date&Clean exposure& Rate  \tabularnewline
      &         &              & EPIC-pn/MOS1/MOS2~(\ks)     &$\rm (counts~s^{-1})$  \tabularnewline
      \hline
      \xmm{}     &  0103861201   &June 7, 2001 & 7.3/11.3/11.6    &$ 0.899\pm0.003$ \tabularnewline
      " &0694580101   &May 29--30, 2012 & 83.9/109.2/110.2  & $2.02\pm0.02$\tabularnewline
      \hline
    \end{tabular*}
    \par\end{centering} {Note-- Count rate for EPIC-pn data in
    $0.3-10$\kev~band.}
\end{table*}

\section{Observation and data reduction}
II~Zw~177 was observed thrice with \xmm{} \citep{2001A&A...365L...1J}. The first two
observations have been studied by various authors
\citep{2005A&A...430..927G,
  2006MNRAS.365..688G,2006MNRAS.368..479G,2008A&A...480..611S,2009A&A...493...55E,
  2011ApJ...727...31A, 2011A&A...530A..42C}. The first observation (ObsID:  0103861201) was
performed starting on 2001 June 7, 04:40 UTC with an exposure time of
13\ks. In this observation, the European Photon Imaging camera (EPIC)
pn \citep{2001A&A...365L..18S} was operated in the full window and two
EPIC-MOS cameras \citep{2001A&A...365L..27T} in the small window mode with
the medium filter. The second observation ({Obs ID: 0103862601) was performed only with
reflection grating spectrometers (RGS1/RGS2) in the spectroscopic mode
with about $3\ks$ each. We performed the third observation in 2012 May (ObsID: 0694580101) for
an exposure of $137\ks$ using the thin filter in the full window
mode (see Table~\ref{obs_log}).  We analyzed EPIC-pn and EPIC-MOS1/MOS2
data from the first and the third (2012) observations.
      
We reprocessed the EPIC-pn and EPIC-MOS data using the Science Analysis System
({\tt SAS}) v13.0.0 \citep{2004ASPC..314..759G} and updated calibration files.  Examination of
light curves extracted above $10\kev$ revealed the presence of flaring
particle background. We created good time interval (gti) files after
excluding the intervals of flaring particle background identified
based on count rate cut-off criteria. We used above criteria by selecting the count
rate cut-off values 2.0, 0.5 in 2012 and 2.4, 0.35 in 2001 observations for EPIC-pn
and EPIC-MOS, respectively. We filtered the event lists to
retain the good events with patterns $\le4$ (EPIC-pn) or $\le12$ (EPIC-MOS)
with arrival times within the good time intervals and generated the
cleaned event lists. This resulted in the useful exposures of $\sim
84\ks$ (EPIC-pn) and $\sim110\ks$ (EPIC-MOS) for the 2012 observation and
$\sim 7\ks$ (EPIC-pn) and $\sim11\ks$ (EPIC-MOS) for the 2001 observation
(see Table~\ref{obs_log}).  We extracted source spectra from circular
regions of radii 55 and 40 arcsec for 2012 and 2001 observations, respectively. We
also extracted the background spectra from off-source circular regions
with radii in the range of 35 to 63 arcsec. We generated the
redistribution matrix and ancillary response files using the SAS tools
{\tt RMFGEN} and {\tt ARFGEN}, respectively. We grouped the spectral
data sets to a minimum counts of 20 per bin and oversampled by a
factor of 5 using the {\tt specgroup} tool.

\begin{table*}
  \small
  \centering 
  \caption{Best-fit spectral model parameters for the two \xmm{}~observations performed in 2012 amd 2001.} \label{pca_ref}
  \begin{tabular}{llcc}
    \hline
    \hline 
    &&\multicolumn{1}{c}{Model 1: Ionized PCA model}\\ 
    \hline
    &    &\multicolumn{2}{c}{\xmm{}}  \\           
    Model&Parameter&0694580101                                  &0103861201\\
    Component && (2012 May)                                     & (2001 June)   \\
    \hline
    Gal. abs.&$N_{H}$ ($10^{20}\rm cm^{-2}$)         & 5.4~(*)                   & 5.4~(*) \tabularnewline
    POWERLAW&$\Gamma$                                &$2.87\pm0.04$              &$2.97_{-0.10}^{+0.08}$\tabularnewline
    &f$_{PL (0.3-2\kev)}$~$~^{a}$                    &$8.4\pm1.0$                &$8.7_{-1.2}^{+2.2}$\tabularnewline  
    &f$_{PL (2-10\kev)}$~$~^{a}$                     &$1.5\pm0.1$                &$1.3_{-0.2}^{+0.1}$\tabularnewline  
    ZXIPCF& $N_{H}$ ($10^{22}{\rm cm^{-2}}$)         &$5.9_{-0.9}^{+15.7}$       &$5.4_{-2.3}^{+2.5}$\tabularnewline
    & $C_{f}$ ($\%$)                                 &$54.3_{-4.4}^{+3.6}$       &$47.0_{-12.5}^{+10.9}$\tabularnewline
    & $\xi$ ($\rm erg~cm~s^{-1}$)                    &$<38.9$                    &$<26.9$\tabularnewline         
    BBODY& $kT_{BB}$ $({\ev}$)                       &$132.3_{-3.2}^{+3.6}$      &$140.9_{-9.3}^{+20.2}$\tabularnewline
    &f$_{BB (0.3-2\kev)}$~$~^{a}$                    &$2.0\pm0.2$                &$1.6_{-0.4}^{+0.3}$\tabularnewline  
    &$\chi^{2}/dof$                                  &$505.6/460$                &$273.4/244$\tabularnewline
    $F_{X}~~^{b}$        &$f_{0.3-2\kev}$            &$3.0$                      &$3.5$  \tabularnewline
    &$f_{2-10\kev}$                                  &$0.95$                     &$0.96$\tabularnewline
    $L_{X}~~^{c}$&$L_{0.3-2\kev}$                    &$4.9$                      &$5.6$ \tabularnewline
    &$L_{2-10\kev}$                                  &$1.6$                      &$1.6$\tabularnewline \hline
    &                       &\multicolumn{1}{c}{Model 2: Blurred reflection model}\\  \hline
    Gal. abs.&$N_{H}$ ($10^{20}\rm cm^{-2}$)         & 5.4~(*)                   & 5.4~(*) \tabularnewline
    NTHCOMP$~^{d}$ & $\Gamma $                       &$2.56_{-0.01}^{+0.02}$     &$2.63_{-0.04}^{+0.05}$ \tabularnewline
    &$kT_{in}$~(\ev)                                 &$10~(*)$                   &$10$~(*) \tabularnewline
    &f$_{NTH (0.3-2\kev)}$~$~^{a}$                   &$1.4_{-0.2}^{+0.3}$        &$1.7\pm1.0$ \tabularnewline  
    &f$_{NTH (2-10\kev)}$~$~^{a}$                    &$0.4\pm0.1$                &$0.5\pm0.2$ \tabularnewline  
   
    REFLIONX & $ A_{Fe}$                             &$1.0_{-0.1}^{+0.2}$        &$1$(*) \tabularnewline
    &$\xi$ ($\rm erg~cm~s^{-1}$)                     &$1161_{-140}^{+131}$       &$1701_{-614}^{+633}$ \tabularnewline
    &$\Gamma$                                        &$2.56_{-0.01}^{+0.02}$ (p$_{t}$) &$2.63_{-0.04}^{+0.05}$ (p$_{t}$)  \tabularnewline
   
    KDBLUR& $ q $                                    &$3.6_{-0.3}^{+0.6}$        &$>3.3$ \tabularnewline
    & $R_{in}$ ($r_{g}$)                             &$<2.97$                    &$2.2_{-0.6}^{+3.0}$ \tabularnewline
    & $R_{out}$ ($r_{g}$)                            &$400$~(*)                  &$400$~(*) \tabularnewline
    & $i$ ($\rm degree$)                             &$<37.6$                    &$43.8_{-28.1}^{+16.1}$ \tabularnewline 
    &f$_{ref (0.3-2\kev)}$~$~^{a}$                   &$3.3_{-0.3}^{+0.2}$        &$3.8\pm1.0$ \tabularnewline  
    &$ \chi^{2}/dof$                                 &$498.2/459$                &$267.8/244$ \tabularnewline
    $ F_{X}~~^{b}$                                   &$f_{0.3-2\kev}$            &$3.0$ &$3.4$ \tabularnewline
    & $f_{2-10\kev}$                                  &$0.89$                    &$0.92$ \tabularnewline
    $ L_{X}~~^{c}$&$L_{0.3-2\kev}$                    &$4.9$                     &$5.6$  \tabularnewline
    &$L_{2-10\kev}$                                   &$1.5$                     &$1.6$ \tabularnewline \hline
    &&\multicolumn{1}{c}{Model 3: Intrinsic disc Comtonised model}\\   \hline
    Gal. abs.&$N_{H}$ ($10^{20}\rm cm^{-2}$)         & 5.4~(*)                   & 5.4~(*) \tabularnewline   
    OPTXAGNF
    &$L/L_{Edd}$                                     &$0.65\pm0.01$              &$0.8_{-0.1}^{+0.8}$\tabularnewline
    &$kT_{e}$ (\kev)                                 &$0.20_{-0.02}^{+0.01}$     &$0.17_{-0.02}^{+0.03}$\tabularnewline
    &$\tau$                                          &$>21.3$                    &$22.1_{-4.6}^{+7.7}$\tabularnewline
    &$r_{corona}~(r_{g})$                            &$3.7_{-0.6}^{+1.4}$        &$<6.8$\tabularnewline
    &$a$                                             &$0.996_{-0.002}^{+P}$      &$0.7_{-0.3}^{+P}$\tabularnewline
    &$f_{pl}$                                        &$0.70\pm0.10$              &$0.6_{-0.2}^{+0.1}$\tabularnewline
    &$\Gamma$                                        &$2.31\pm0.05$              &$2.55_{-0.10}^{+0.09}$\tabularnewline
    &f$_{OPTX (0.3-2\kev)}$~$~^{a}$                  &$4.68_{-0.05}^{+0.04}$     &$5.6\pm0.2$\tabularnewline  
    &f$_{OPTX (2-10\kev)}$~$~^{a}$                   &$0.97_{-0.02}^{+0.03}$     &$0.97\pm0.05$\tabularnewline  

    &$\chi^{2}/dof$                                  &$502.4/460$                &$275.0/244$\tabularnewline
    $F_{X}~~^{b}$&$f_{0.3-2\kev}$                     &$3.0$                      &$3.4$\tabularnewline
    &$f_{2-10\kev}$                                   &$0.96$                     &$0.97$\tabularnewline
    $L_{X}~~^{c}$&$L_{0.3-2\kev}$                     &$4.8$                      &$5.6$ \tabularnewline
    &$L_{2-10\kev}$                                   &$1.6$                      &$1.7$\tabularnewline
    \hline

  \end{tabular}
  {~~~~~~~~~~~~~~~~~~~~~~~~~~~~~~~~~~~~~~~~~~~~~~~~~~~~~~~~~~~~~~~~~~~~~~~~~~~~~~~~~~~~~~~~~~~~~Notes-- (a) f$_{PL}$, f$_{NTH}$, f$_{ref}$ and f$_{OPTX}$ represent fluxes to respective model component and flux is measured in units $10^{-12}{\rm~ergs~cm^{-2}~s^{-1}}$; (b) This is observed flux in units $10^{-12}{\rm~ergs~cm^{-2}~s^{-1}}$ of a given energy band.; (c) Luminosity in units $10^{43}{\rm~ergs~s^{-1}}$; (d) Electron temperature is fixed to 100\kev. P stands for pegged to hard limit.  'p$_{t}$' stands for parameter tied. Fixed parameters are indicated by an asterisk.}
\end{table*}

\begin{table*}
  \small
   \centering 
  \caption{Best-fit spectral model parameters for the two \xmm{}~observations performed in 2012 amd 2001.} \label{lp_ref}
  \begin{tabular}{llcc} \hline
 &                       &\multicolumn{1}{c}{Model : reflection model with RELCONV\_LP}\\ 
  \hline
  Model&Parameter&0694580101                                  & 0103861201\\
    Component && (2012 May)                                     & (2001 June)   \\
 
    \hline
       Gal. abs.&$N_{H}$ ($10^{20}\rm cm^{-2}$)         & 5.4~(*)                & 5.4~(*) \tabularnewline
    NTHCOMP$~^{d}$ & $\Gamma $                       &$2.56\pm0.02$              &$2.6\pm0.1$  \tabularnewline
    &$kT_{in}$~(\ev)                                 &$10~(*)$                   &$10$~(*) \tabularnewline

    &f$_{NTH (0.3-2\kev)}$~$~^{a}$                   &$1.4_{-0.4}^{+0.3}$        &$2.1\pm1.0$ \tabularnewline  
    &f$_{NTH (2-10\kev)}$~$~^{a}$                    &$0.4\pm0.1$                &$0.6\pm0.2$ \tabularnewline 
    REFLIONX & $ A_{Fe}$                             &$1.0_{-0.1}^{+0.2}$        &$1$(f)  \tabularnewline
    &$\xi$ ($\rm erg~cm~s^{-1}$)                     &$1135.2_{-107.4}^{+184.1}$  &$1323_{-379}^{+771}$ \tabularnewline
    &$\Gamma$                                        &$2.56\pm0.02$ (p$_{t}$)          &$2.6\pm0.1$ (p$_{t}$) \tabularnewline
    RELCONV\_LP& $ h $ ($r_{g}$)                     &$3.3_{-1.4}^{+0.4}$        &$<4.7$\tabularnewline
    &$\Gamma$                                        &$2.56\pm0.02$ (p$_{t}$)          &$2.6\pm0.1$ (p$_{t}$)    \tabularnewline
    & $R_{in}$ ($r_{g}$)                             &$<3.1$                     &$<4.1$ \tabularnewline
    & a                                              &$>0.7$                     &$>0.3$ \tabularnewline
    & $R_{out}$ ($r_{g}$)                            &$400$~(*)                  &$400$~(*) \tabularnewline
    & $i$ ($\rm degree$)                             &$18.6_{-4.3}^{+9.2}$       &$<44.3$ \tabularnewline 

 &f$_{ref (0.3-2\kev)}$~$~^{a}$                      &$3.4_{-0.3}^{+0.4}$        &$3.4\pm1.0$ \tabularnewline     
    &$ \chi^{2}/dof$                                 &$500.2/458$                &$271.9/243$ \tabularnewline
    $ F_{X}~~^{b}$        &$f_{0.3-2\kev}$            &$3.0$                     &$3.5$ \tabularnewline
    & $f_{2-10\kev}$                                  &$0.89$                     &$0.93$ \tabularnewline
    $ L_{X}~~^{c}$&$L_{0.3-2\kev}$                    &$4.9$                      &$5.6$  \tabularnewline
    &$L_{2-10\kev}$                                   &$1.5$                      &$1.6$ \tabularnewline \hline
    \hline

  \end{tabular}
  {{~~~~~~~~~~~~~~~~~~~~~~~~~~~~~~~~~~~~~~~~~~~~~~~~~~~~~~~~~~~~~~~~~~~~~~~~~~~~~~~~~~~~~~~~~~~~~Notes-- (a) f$_{NTH}$ and f$_{ref}$ represent fluxes to respective model component and flux is measured in units of $10^{-12}{\rm~ergs~cm^{-2}~s^{-1}}$; (b) This is observed flux in units of $10^{-12}{\rm~ergs~cm^{-2}~s^{-1}}$ for a given energy band.; (c) Luminosity in units of $10^{43}{\rm~ergs~s^{-1}}$; (d) Electron temperature is fixed to 100\kev. 'p$_{t}$' stands for parameter tied. Fixed parameters are indicated by an asterisk.}}
\end{table*}

\begin{figure*}
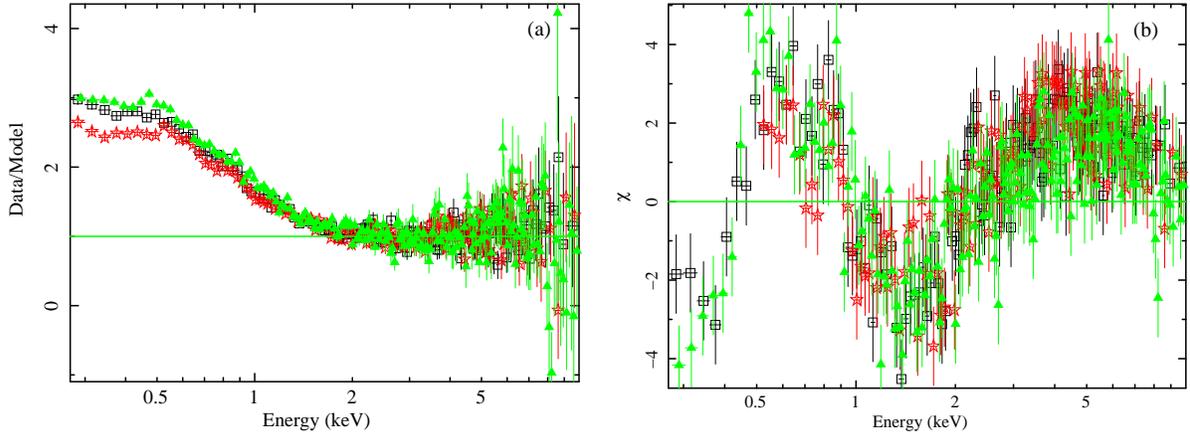

  \centering
  \includegraphics[height=8cm,angle=-90]{fig1a.ps}
  \includegraphics[height=8cm,angle=-90]{fig1b.ps}
  \caption{2012 long observation: (a) The data-to-model ratio based on the simple absorbed
    POWERLAW model fitted to the EPIC-MOS1 (open squares), EPIC-MOS2 (asterisks)
    and EPIC-pn (filled triangles) in the the $2-10\kev$ band and
    extrapolated to lower energies (b) Residuals in terms
      of $\chi=\rm~(Data-Model)/\sigma$ for fitted absorbed POWERLAW
      model in $0.3-10\kev$ band. }
  \label{pca_mod}
\end{figure*}

\begin{figure*}
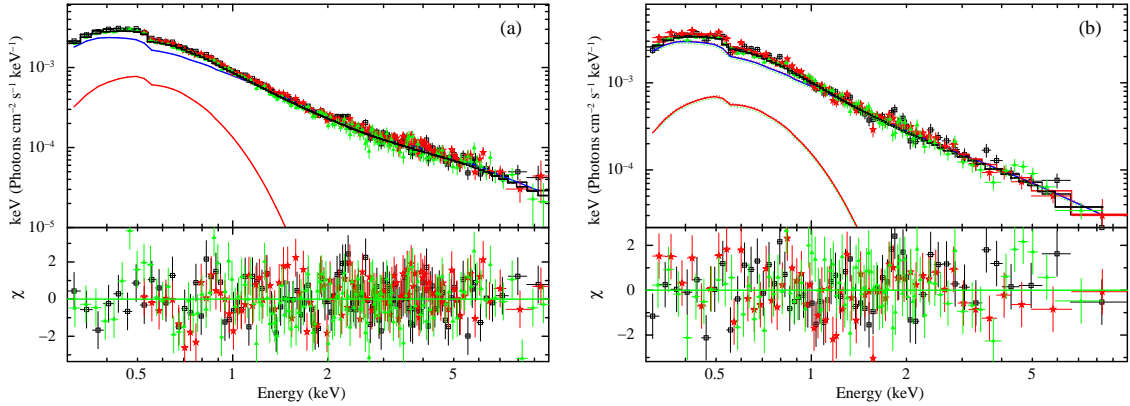

  \includegraphics[scale=0.3,angle=-90]{fig2a.eps}
  \includegraphics[scale=0.3,angle=-90]{fig2b.eps}
  \caption{ Results of spectral fits to the 2012 (left) and 2001 (right) spectral data, the best-fitting ionized PCA
    (WABS$\times$ZXIPCF$\times$(POWERLAW+ZBBODY)), and the deviations of the observed data from the
    model. The symbols are the same as used in Fig.~\ref{pca_mod}.}
  \label{pca_fit_spec}
\end{figure*}

\section{X-ray Spectral modeling}
We used {\tt XSPEC} v12.8 \citep{1996ASPC..101...17A} to analyze the spectral data sets and
employed the $\chi^2$ statistics to find the best-fit models.  The
errors on each best-fitting parameter reflect $90\%$ confidence level
unless otherwise stated.

We begin with the spectral analysis of 2012 data. First, we compared
the EPIC-pn, EPIC-MOS1 and EPIC-MOS2 spectral data by fitting the $2-10\kev$
band by a simple model CONS--\\TANT$\times$WABS$\times$POWERLAW.  Here the
constant has been introduced to take into account any difference
in  flux normalization  between the instruments. 
 We fixed the Galactic absorption column density
$N_{H}=5.4\times10^{20}$~$\rm~cm^{-2}$
\citep{1990ARA&A..28..215D}. The fit resulted in
$\chi^{2}=293.5$ for 309 degree of freedom (dof) with $\Gamma=2.35\pm{0.05}$. We then
extrapolated the model down to $0.3\kev$ and plotted the data-to-model
ratio in Figu~\ref{pca_mod}~(a). We noticed that the EPIC-MOS2 data do not
agree well below $0.5\kev$ with the EPIC-pn or EPIC-MOS1 data.  We
therefore excluded EPIC-MOS2 data below $0.5\kev$ from our spectral
analysis.  In addition, the ratio plot clearly shows strong soft X-ray
excess emission below $2\kev$.
 
The soft X-ray component has been observed in many Seyfert 1/QSO AGN
(e.g., \citealt{1985ApJ...297..633S,1985MNRAS.217..105A,1996A&A...305...53B,
  1999MNRAS.309..113V,2006MNRAS.365.1067C}).
Though the origin of the soft X-ray excess is not clearly understood,
the component can be modeled by three physically motivated models --
partially ionized absorption,
blurred reflection from a partially ionised accretion disc and the
thermal Comptonization in an optically thick, cool plasma
\citep{1985MNRAS.217..105A,1999ApJS..125..317L,2006MNRAS.365.1067C,
  2007ApJ...671.1284D, 2007MNRAS.374..150S,2012MNRAS.420.1848D}.  Here
we use these models and test their validity using the UV/X-ray
variability. We note that fitting the absorbed powerlaw model to the
full $0.3-10\kev$ band provided a poor fit
($\chi^{2}/dof=1590.1/465$) and resulted in the curvature in the
$1.5-5\kev$ band and excess emission in the $0.5-1\kev$ emission as shown
in Fig.~\ref{pca_mod}~(b). Such features may be caused by complex partial
covering absorption (PCA) which we test first.

\subsection{Ionized partial covering model}

Multiplying the simple absorbed powerlaw model with a
  neutral partial covering absorption (ZPCFABS) model improved the fit
  ($\chi^2/dof=887.2/463$) with $N_H\sim 1.4\times10^{23}~cm^{-2}$ and covering fraction $C_{f}$
  $\sim 59.4\%$. However the fit is statistically unacceptable. We also
  tested the possible presence of partially covering ionised
  absorption along the line of sight to the nuclear source.  We used
  the ZXIPCF model which is an XSTAR based absorption model
  constructed with illuminating continuum slope of $\Gamma=2.2$ and
  turbulent velocity of $200{\rm~km~s^{-1}}$
  \citep{2008MNRAS.385L.108R}.  Replacing the neutral PCA component
  with the ionised PCA model ZXIPCF improved the fit
  ($\Delta\chi^2=-9.4$ for one additional parameter. Using a second ZXIPCF
  component did not improve the fit further. Thus, complex
  absorption cannot describe the observed data. Indeed, adding a simple
  blackbody i.e., the model WABS$\times$ZXIPCF$\times$(POWERLAW+ZBBODY) 
  improved the fit considerable ($\chi^2/dof = 505.6/460$). 
 The blackbody temperature is  $kT_{BB}
  \sim 132\ev$ which is typical of AGN with strong soft X-ray excess.  
Although this model does not fit the data in the sense that it 
requires an adhoc blackbody component (see Fig.~\ref{pca_fit_spec}~(a)), for completeness 
the best-fit parameters are listed   in Table~\ref{pca_ref}.

\begin{figure*}
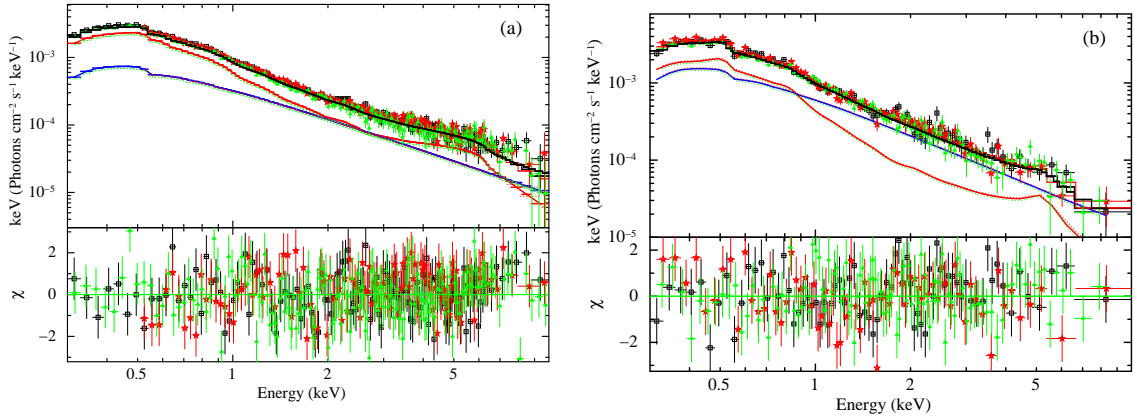


  \includegraphics[scale=0.3,angle=-90]{fig3a.eps}
  \includegraphics[scale=0.3,angle=-90]{fig3b.eps}
   \caption{ Results of spectral fits to the 2012 (left) and 2001 (right) spectral data, the best-fitting blurred reflection
    (WABS$\times$(NTHCOMP+KDBLUR*REFLIONX) and the deviations of the observed data from the
    model. The symbols are the same as used in Fig.~\ref{pca_mod}.}
  \label{ref_fit_spec}
\end{figure*}

\begin{figure*}
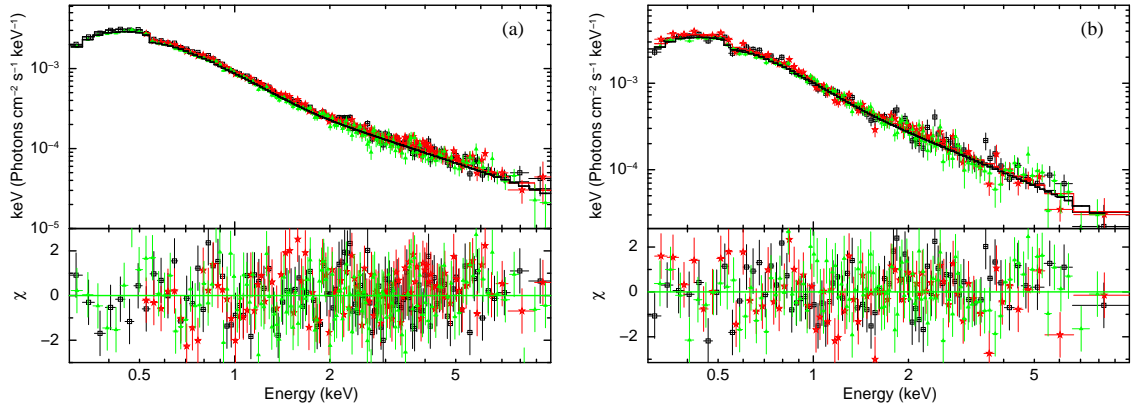

  \includegraphics[scale=0.3,angle=-90]{fig4a.eps}
  \includegraphics[scale=0.3,angle=-90]{fig4b.eps}
  \caption{ Results of spectral fits to the 2012 (left) and 2001 (right) spectral data, the best-fitting intrinsic disc model
    (WABS$\times$OPTXAGNF), and the deviations of the observed data from the
    model. The symbols are the same as used in Fig.~\ref{pca_mod}.}
  \label{disc_fit_spec}
\end{figure*}

\begin{figure*}
  \includegraphics[scale=0.3,angle=-90]{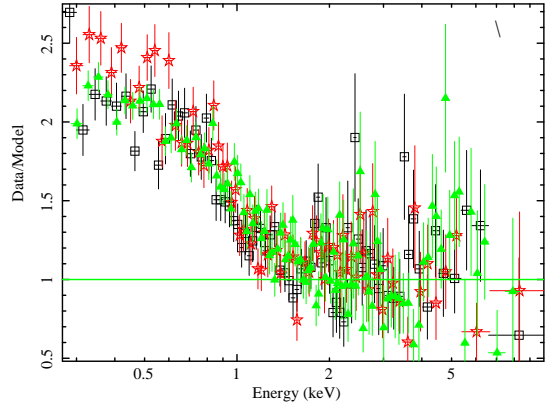}
  \caption{ Ratio of observed 2001 data and the best-fit $2-10\kev$ absorbed
    powerlaw model in the $0.3-10\kev$ band. The symbols are the same as used in Fig.~\ref{pca_mod}.}
  \label{excess_fit_spec}
\end{figure*}

\subsection{The blurred reflection model}
The soft X-ray excess emission can be described by reflection from a
relativistic photoionized accretion disc. In the ``lamppost'' geometry, a
hot compact plasma, above the accretion disc, emits the primary
emission which illuminates the surface of disc material
\citep{1991A&A...247...25M}. The X-ray illumination is stronger in the
inner regions of the disc due to the strong gravity of the SMBH (e.g., \citealt{2004MNRAS.349.1435M}).
The emissivity of the reflected emission is often expressed as
$\epsilon\propto~r^{-q}$, where $r$ is the distance from the center
and $q$ is the emissivity index.  The soft X-ray excess can arise due
to the numerous emission lines from the photoionised disc and Thomson
scattering of the illuminating powerlaw emission. 
 We therefore modeled the soft X-ray excess emission using the REFLIONX model which
characterizes the reflected emission from partially ionized accretion
disc \citep{2005MNRAS.358..211R}.  The parameters of REFLIONX are iron abundance
relative to solar ($A_{Fe}$), ionization parameter $\xi=L/nr^{2}$ ( where $L$ is
the source luminosity, $n$ is hydrogen density and r represents the distance between
source and disc) and photon index which is same as the  
illuminating powerlaw photon index. Thus, we tied REFLIONX $\Gamma$ to POWERLAW $\Gamma$
required for disc illumination. }The model WABS$\times$(POWERLAW+REFLIONX) improved the fit
($\chi^{2}/dof=963.3/462$) compared to the absorbed powerlaw. The residuals
in above modeling could be smoothed by the gravitational and
relativistic effects close to the SMBH \citep{2000PASP..112.1145F,2009Natur.459..540F}.
We blurred the ionized reflected emission by the convolution model KDBLUR to obtain
the relativistically broaden ionized reflection from the inner regions of the accretion disc.
The parameters of KDBLUR are emissivity index ($q$), inner and outer radii of the disc
(R$_{in}$ and $R_{out}$) and inclination ($i$) angle between observer's line of sight and normal to the disc.
The WABS$\times$(POWERLAW+KDBLUR*REFLIONX) resulted in a satisfactory fit
($\chi^{2}/dof=497.2/459$).  To be more
physically consistent, we replaced the powerlaw by a thermal 
Comptonization model NTHCOMP \citep{1996MNRAS.283..193Z}. 
In this model, we fixed the electron
temperature to a high value of $100\kev$ and soft disc photon temperature to
$10\ev$. The model WABS$\times$(NTHCOMP+KDBLUR*REFLIONX) resulted
in the similar $\chi^{2}/dof=498.2/459$ as expected. We did not find any significant
residuals and consider this as the best-fit blurred reflection
model. The spectral data, the best-fit model, and the residuals are
shown in Fig.~\ref{ref_fit_spec}~(a) and the best-fitting parameters are listed in
Table~\ref{pca_ref}. We further replaced KDBLUR by RELCONV\_LP, which also
convolves with the entire reflected X-ray spectrum to blur the emission features and provides lamppost
height of the X-ray source. The important model parameters are height of the compact corona $h ~(r_{g})$, inner disc radius, spin, outer disc radius, inclination and illuminatiing powerlaw index.
The best-fit parameters of WABS$\times$(NTHCOMP+RELCONV\_LP*REFLIONX) are listed in Table~\ref{lp_ref}.

\begin{figure}
  \centering
 \includegraphics[height=9cm,angle=-90]{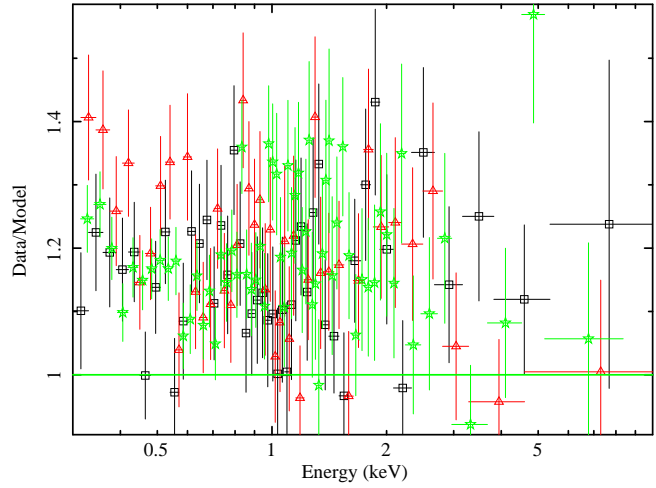} \\
  \caption{ The ratio (data/model) of the 2001 EPIC spectral data sets and the best-fit NTHCOMP+DISKBB model
    of the 2012 EPIC data sets.}
  \label{difference_model}
\end{figure}

\begin{figure}
  \centering
  \includegraphics[height=8cm, angle=0]{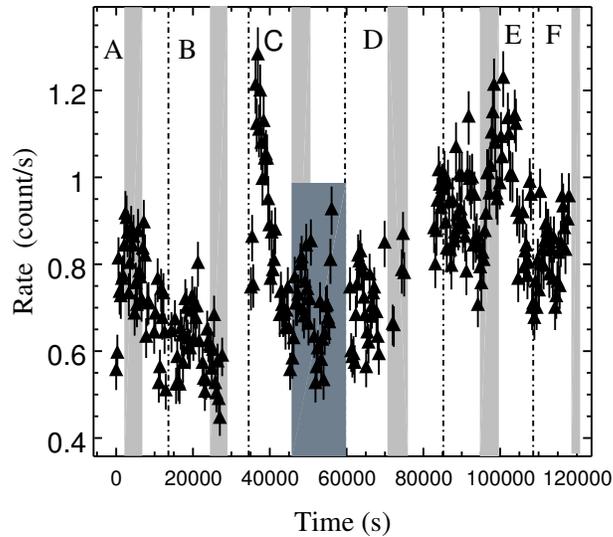}
  \caption{ The selection for time slices from 2012 May EPIC-pn
  lightcurve for time resolved spectroscopy. Dash-dotted lines show time slices of about 20ks each (A,B,C,D,E,F).  The shaded light gray time slices are the exposures exactly similar to the observed U band emission. The shaded dark gray time slice ($45-60$ ks) where flare is excluded from C interval in the analysis.}
  \label{time_select_curve}
\end{figure}

\subsection{The intrinsic disc Comptonization model}
The strong soft X-ray excess emission can also be modeled by thermal
Comptonization in a cool, optically thick plasma \citep[see
e.g.,][]{2007ApJ...671.1284D}. If the gravitational energy released is
not completely thermalised, then it is possible that a fraction of the
energy  dissipated in the inner disc elevates its
temperature such that it acts as an optically thick thermal
Comptonizing medium. The intrinsic disc Comptonization model (OPTXAGNF)
accounts for the thermal Comptonization in the inner disc material
\citep{2012MNRAS.420.1848D}.
This model is an energetically consistent model in which gravitational
energy is distributed among different parts of the accretion disc.
This model itself takes care of colour
temperature correction at each radius to partly contribute to the soft excess.
The inner disc emission gets Compton upscatters from thermal Comptonization medium,
$r_{corona}$ to r$_{ISCO}$,  which consists of an
optically thick cool ($kT_{e}\sim0.2$\kev) electron plasma.  The thermal Comptonization in
the disc plasma is responsible for the soft X-ray excess emission. A
fraction of total gravitational energy is also dissipated into a hot
corona below $r_{corona}$ which is responsible for the hard X-ray
emission extending to a few $100\kev$. Thus, the OPTXAGNF model includes the
three components -- thermal emission from a standard disc, optically thick thermal
Comptonization in the inner disc and optically thin thermal Comptonization in a hot
corona below $r_{corona}$. This model provides important information on black hole spin ($a$), accretion
rate relative to Eddington rate ($L/L_{Edd}$) and properties of optically thick
cool plasma (elctron temperature $kT_{e}$ and optical thickness $\tau$).

To fit the soft X-ray excess and the high energy continuum, we used
the model WABS$\times$OPTXAGNF. We fixed the black hole mass to
$2\times10^{6}~M_{\odot}$  \citep{2007ApJ...667..131G, 
  2011ApJ...727...31A}, distance to $364\mpc$ ($z=0.081$)
and outer radius to $400r_{g}$. 
 The residuals, obtained after fitting the model,
did not show any significant features with a 
$\chi^{2}/dof=502.4/460$.  The observed data, the model and the
residuals are shown in Fig.~\ref{disc_fit_spec}~(a) and the best-fit parameters
are listed in Table \ref{pca_ref}. There are no obvious narrow emission
line in the Fe-K regime which may arise from 
a neutral reflection component from a distant material such
as a putative torus. To examine this further, we included a narrow Gaussian line
with the width fixed at $\sigma=10\ev$, line energy fixed to $6.4\kev$ and obtained a $90\%$ upper limit
on the equivalent width to be 0.08\kev.

Since the model includes the emission from the outer disc, it can
also predict the UV emission from such a system. This can be compared with
the data from the Optical Monitor (OM). The reduction and reprocessing of the UV data
  measured with the OM is described in detail in section 5. We
  used the time-averaged count rate in the U filter to construct the U
  band spectrum usable in XSPEC. We also used the latest response for
  this filter \footnote{http://xmm2.esac.esa.int/external/xmm\_sw\_cal/calib/om\_files.shtml}.
  We obtained the flux density in the U band to be $7.49\pm0.03\times10^{-16}$\efluxA.
  We estimated the predicted U band flux by extending the best fit X-ray model. For this,
we increased the outer radius to $10^5 r_g$ and used a Galactic reddening factor
$E(B-V)=0.0681$ \citep{2011ApJ...737..103S}. The estimated flux density was found to
be $0.810\pm0.001\times10^{-16}$\efluxA. Clearly the observed UV flux 
is significantly higher than the predicted value indicating that the UV emission is not due
to the outer disc intrinsic emission but perhaps due to X-ray reprocessing. In this
analysis, we have fixed the black hole mass to  $M_{BH}= 2\times10^{6}~M_{\odot}$
estimated from the $M_{BH}$-line width- luminosity relation described in
  \citet{2007ApJ...667..131G, 2007ApJ...670...92G}  using FWHM
  width of $H_{\alpha}$ line. However, in addition, if we allow the mass 
  to vary, the predicted UV flux matches with the observed one
  for $M_{BH}=3.8_{-1.9}^{+1.5}\times10^{7}~M_{\odot}$. The best-fit resulted in $\chi^{2}/dof=516.5/460$.  The best-fit parameters of {\tt TBABS$\times$REDDEN$\times$OPTXAGNF} are as follows: $L/L_{Edd}=<0.32$, $kT_{e}=0.19\pm0.01$~\kev, $\tau=19.9_{-1.4}^{+1.7}$, $r_{corona}=63.8_{-31.7}^{+P}~r_{g}$ (P stands for pegged to hard limit), $a<0.93$, $f_{pl}=0.4\pm0.1$ and $\Gamma=2.35_{-0.05}^{+0.04}$ . Since there may be uncertainties in the black
  hole mass measurement this can be a viable option.

 \section{Long Term X-ray Spectral Variability}

To study long term spectral variability, we next analyzed and  fitted
the 2001 spectral data.  As before, we compared the
EPIC-pn, EPIC-MOS1 and EPIC-MOS2 datasets by fitting the WABS$\times$POWERLAW
model in the $2-10\kev$ band and comparing the data-to-model ratios in
the full $0.3-10\kev$ band after extrapolating the powerlaw model to
the low energies.  We did not find any significant calibration issues
among the data sets and as before we detected strong soft X-ray excess
emission (see Fig.~\ref{excess_fit_spec}). We then modeled the
broadband spectrum in the $0.3-10\kev$ using the same three model as used for
the 2012 data.  These fits
resulted in $\chi^{2}/dof=273.4/244$, $261.9/243$ and $322.8/245$
for ionized partial covering absorption, blurred reflection and
intrinsic disc models, respectively. The best-fit models, data and the
residuals are shown in Fig.~\ref{pca_fit_spec}~(b),~\ref{ref_fit_spec}(b)
and ~\ref{disc_fit_spec}(b), and the best-fit parameters are listed in
Table~\ref{pca_ref}. As before, we also replaced KDBLUR with
RELCONV\_LP to estimate the height implied for the X-ray source in the lamppost
illumination model. The best-fit parameters of WABS$\times$(NTHCOMP+RELCONV\_LP*REFLIONX)
are listed in Table~\ref{lp_ref}.

 To further investigate the long term variability, we fitted both 2001 and 2012 observation by NTHCOMP+DISKBB model. We used NTHCOMP as a powerlaw component and DISKBB to describe the soft excess component. The soft excess flux obtained from DISKBB in the $0.3-2$ \kev~band and the NTHCOMP flux in the $2-10$ \kev~band remained nearly constant. The NTHCOMP flux in the $0.3-2$\kev~band was about 16\% higher and the photon index $\Gamma$ slightly steeper in the 2001 observation as compared to 2012 observation indicating that the powerlaw component flattened at lower flux in 2012 (see Table ~\ref{diff_spec}). We applied the best-fit simple model (NTHCOMP+DISKBB) of 2012 observation to 2001 observation and show the ratio of 2001 data and 2012 model in Fig.~\ref{difference_model}.  The marginal excess is due to slightly steeper powerlaw component at higher flux in 2001. Such spectral variability i.e., steeper spectrum at high flux, is common in Seyfert 1 galaxies (e.g., \citealt{1998ApJ...505..594N, 2001ApJ...548..694V, 2001ApJ...551..186Z, 2002ApJ...564..162R, 2011MNRAS.415.1895E}). 
	 
\begin{table*}
  \noindent
  \caption{Fluxes for soft X-ray excess and powerlaw in different~bands modeled with DISKBB+NTHCOMP} \label{diff_spec}
  \begin{tabular}{lccccc}
    \hline 
    Observation   &$\Gamma$   &$kT_{multicbb}~(\ev)$     &$f_{SE}$~($0.3-2$\kev)   &$f_{NTH}$ ($0.3-2$\kev)  &$f_{NTH}$ ($2-10$\kev)\tabularnewline
    \hline
    2001    &$2.62\pm0.07$ &$154.0_{-5.7}^{+5.5}$  &$2.0\pm0.3$   &$3.4\pm0.3$   &$0.95\pm0.05$ \tabularnewline
    2012    &$2.49\pm0.03$ &$156.9\pm2.1$   &$2.0\pm0.1$            &$2.6\pm0.1$           &$0.92\pm0.02$ \tabularnewline
    \hline
  \end{tabular}
  {~~~~~~~~~~~~~~~~~~~~~~~~~~~~~~~~~~~~~~~~~~~~~~~~~~~~~~~~~~~~~~~~~~~~~~~~~~~~~~~~~~~~~~~~~~~~~~~~~~~~~~~~~~~~~~~~~~~~~~~~~~~~~~~~~~~~~~~~~~~~~~~~~~Note--Flux is measured in units of $10^{-12}$ \funit.}
\end{table*}

\begin{table*}
  \noindent \begin{centering}
	  \caption{Best-fit spectral model parameters for the time-selected EPIC-pn spectra extracted based on OM exposures, and also similarly for time slices of flares.} \label{pldbb1}
    \begin{tabular}{lccccccc}
      \hline
      \hline 
      Exposure &\multicolumn{3}{c}{ POWERLAW}  & \multicolumn{3}{c}{Soft X-ray excess}    \tabularnewline
      interval (ks) &$\Gamma$ &n$_{pl}$~($10^{-4}$) &$f_{2-10\kev}$~($10^{-13}$) &kT$_{in}$(\ev)&n$_{multicbb}$~($10^{2}$) &$f_{0.3-2\kev}$~($10^{-12}$) &$\chi^{2}/dof$ \tabularnewline
      \hline
      0-14~(\rm A)&$2.5\pm0.2$                &$6.4\pm1.0$   &$7.9\pm1.1$  &$164.4_{-9.7}^{+9.5}$ &$2.5_{-0.8}^{+0.9}$ &$1.95_{-0.5}^{+0.4}$&69.0/76\tabularnewline

      14-35~(\rm B)&$2.5\pm0.2$ &$6.1\pm0.7$    &$7.2_{-0.8}^{+0.9}$  &$159.1_{-11.1}^{+10.7}$  &$2.0_{-0.7}^{+0.8}$    &$1.3_{-0.4}^{+0.3}$&66.2/86\tabularnewline
      35-60~(\rm C)&$2.7\pm0.1$ &$7.9\pm0.7$   &$7.8\pm0.7$          &$156.7_{-7.9}^{+8.1}$    &$2.7_{-0.8}^{+0.9}$     &$1.7_{-0.4}^{+0.3}$&135.7/111\tabularnewline
      60-85~(\rm D)&$2.5\pm0.2$ &$6.3\pm0.8$   &$7.5_{-0.9}^{+1.0}$  &$157.5_{-8.4}^{+8.0}$    &$3.0_{-0.8}^{+0.9}$     &$1.9\pm0.4$&99.9/91\tabularnewline
      85-109~(\rm E)&$2.6\pm0.1$ &$9.0\pm0.7$  &$9.85_{-0.7}^{+0.8}$  &$151.2_{-5.5}^{+5.3}$    &$4.9_{-0.9}^{+1.0}$     &$2.5_{-0.4}^{+0.3}$&128.0/120\tabularnewline
      109-126~(\rm F)&$2.5\pm0.2 $&$8.4\pm1.1$&$9.9_{-1.1}^{+1.2}$&$162.7_{-10.4}^{+9.9}$ &$2.9_{-0.9}^{+1.0}$   &$2.1\pm0.5$&75.9/84\tabularnewline
      35-45~(\rm C flare)&$2.5\pm0.2$ &$8.3\pm1.0$   &$10.1\pm1.3$  &$151.5_{-9.8}^{+9.0}$    &$4.4_{-1.2}^{+1.5}$     &$2.3\pm0.5$  &90.0/84\tabularnewline
      95-105~(\rm E flare)&$2.4\pm0.2$ &$8.6\pm1.0$  &$12.5_{-1.3}^{+1.4}$  &$150.3_{-7.7}^{+7.3}$    &$6.5_{-1.3}^{+1.6}$     &$3.3_{-0.5}^{+0.4}$&102.2/86\tabularnewline

      \hline
    \end{tabular}
    \par\end{centering}
    {~~~~~~~~Note-- $n$ represents the normalization of respective components and  flux is measured in units of \funit~for EPIC-pn data. The fluxes of both soft excess and powerlaw components were measured from DISKBB and POWERLAW model convolving with CFLUX in the $0.3-2$\kev ~and $2-10$\kev~ bands, respectively. We also modified above model by Galactic absorption while fitting.}
\end{table*}

  \begin{figure*}
    \centering
    \includegraphics[scale=0.6]{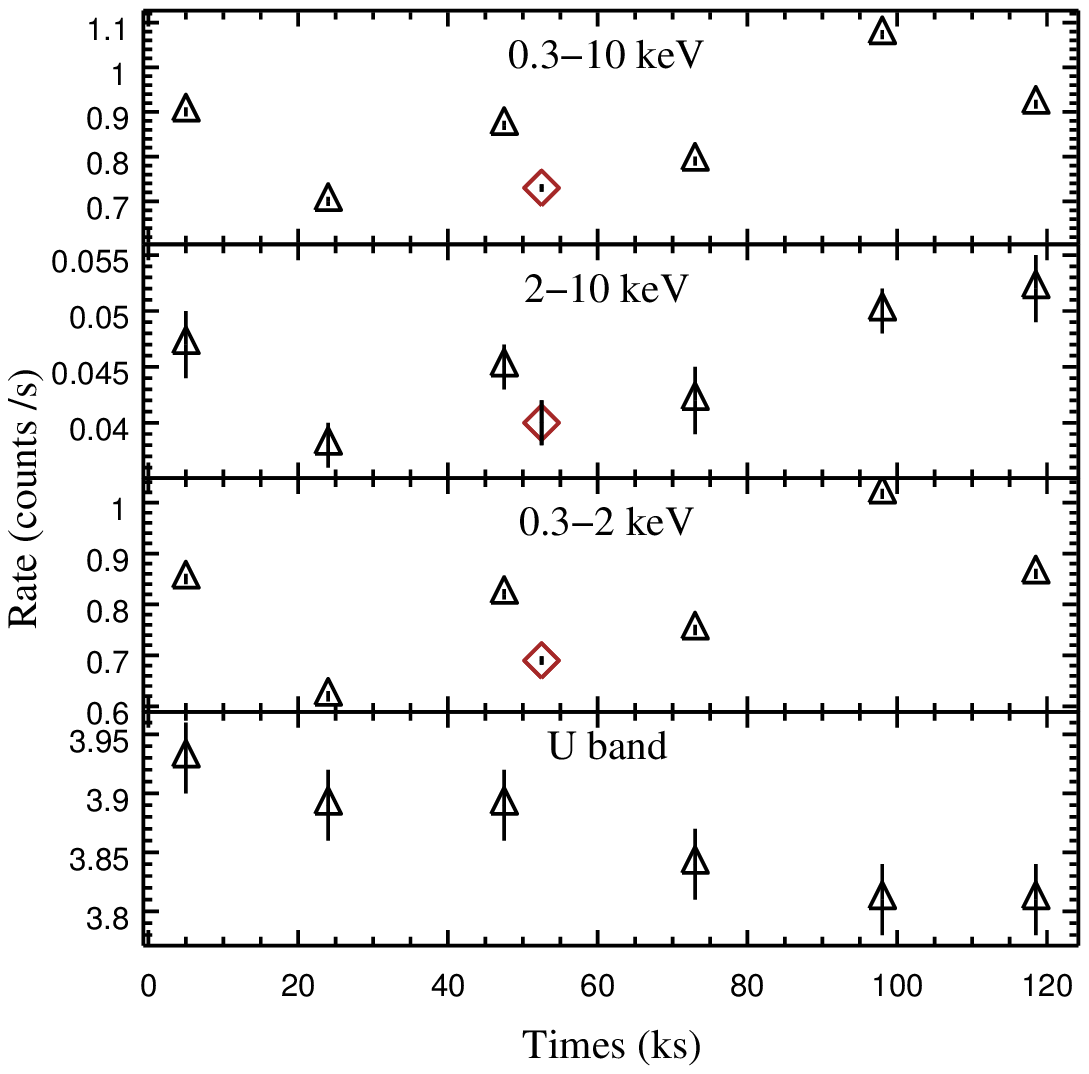}
    \includegraphics[height=8cm, angle=0]{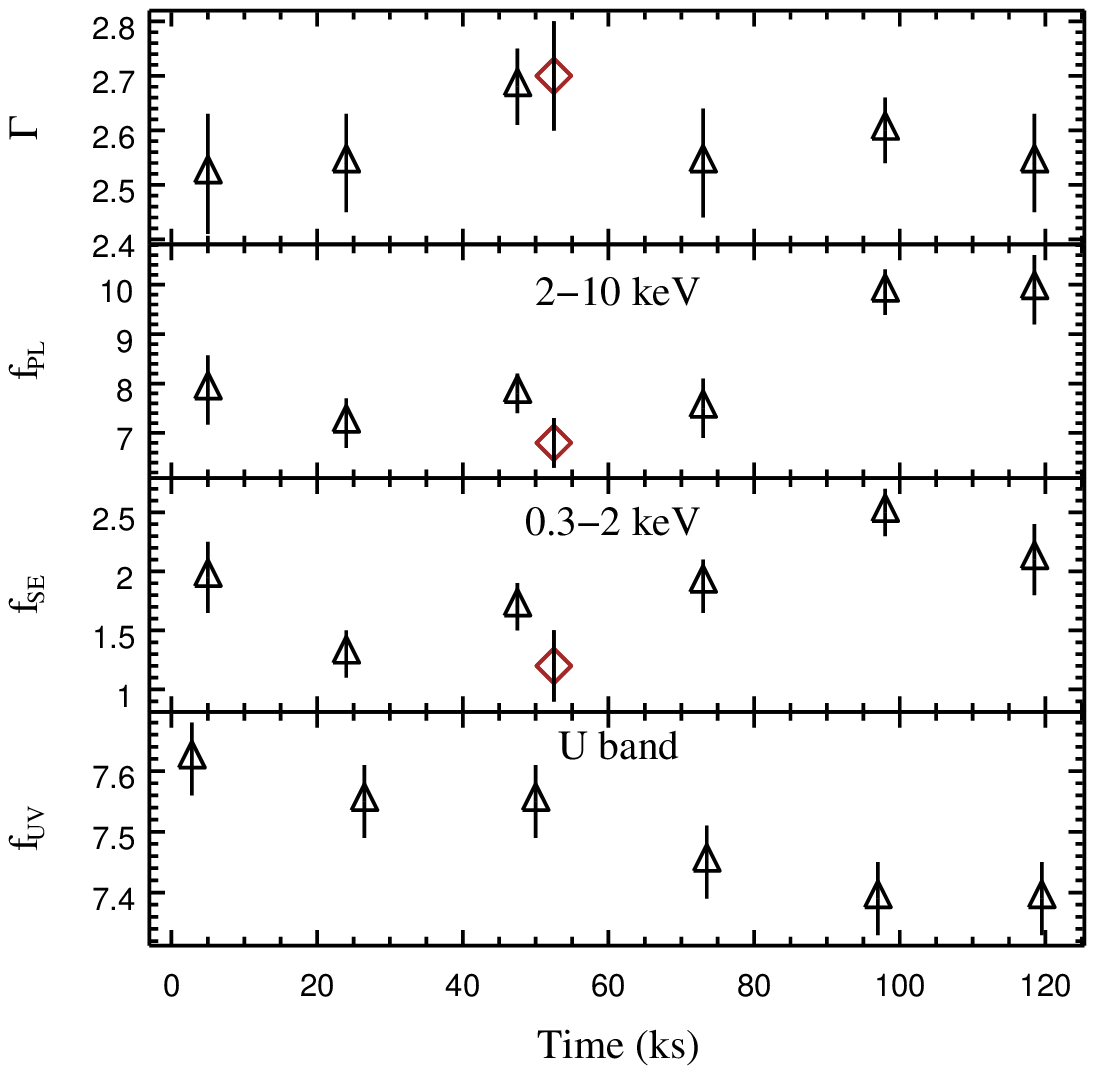}
    \caption{ Left panel: The model independent evolution of different
      spectral components constructed using mean sub-exposures.  Right panel:
      The model dependent evolution of different spectral components
      constructed using mean sub-exposures where powerlaw and soft
      excess flux are plotted in units of $10^{-13}$ and
      $10^{-12}$\funit~,respectively. The UV flux density is measured in units
      of $10^{-16}$~\efluxA. The diamond symbol indicates the C interval where X-ray
      flare is excluded from the analysis. The error bars are plotted with one sigma
      error to the values.}
    \label{time_sel_results}
  \end{figure*}

   \begin{figure*}
    \centering
    \includegraphics[scale=0.6]{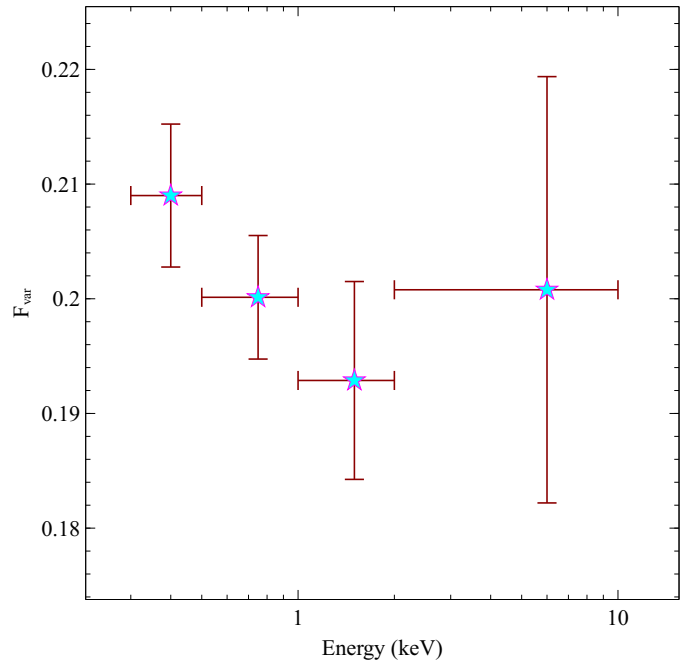}
    \caption{ Fractional variability amplitude F$_{var}$ as a function of energy derived from background corrected EPIC-pn lightcurve with 500s bins. }
    \label{rms_spec}
  \end{figure*}

\section{ X-ray and UV variability during 2012 observation}

The long observation of 2012 allows us to study the short term variability 
of the X-rays and the UV emission from the source. 
We constructed the lightcurve for the UV emission at
$3440$\text{~\AA}  measured with the Optical Monitor
\citep[{OM}:][]{2001A&A...365L..36M}.  First we reprocessed the OM
data with the OMICHAIN tool. In the total span of $137\ks$ exposure,
we obtained 6 OM exposures (IDs: 025, 026, 027, 028, 029 and 030),
$\sim 4\ks$ each, of U filter centered at $3440$\text{~\AA}. We
obtained the source count rates from the source list files
generated by the OMICHAIN tool. We also verified these corrected count
rates by performing interactive photometry of the source. We selected
source region with a radius of 6 pixels and an annular area for
background region with inner radius of 8 pixels and outer radius of 13
pixels and obtained the background subtracted count rates from each OM
exposure. We used a conversion
factor\footnote{$http://xmm.esac.esa.int/external/xmm\_user\_support/documentation/$}
of 1.94$\times10^{-16}$\efluxA~to convert the count rates to flux densities
at $3440$\text{~\AA} and constructed the OM UV lightcurve.
We also derived the UV lightcurves of nearby four sources 
and compared to the UV lightcurve of II~Zw~177. We found that the UV emission from
the nearby sources did not vary as the UV emission of this AGN. Hence we concluded that
the UV variability of II~Zw~177 is intrinsically associated with the AGN.

Further, we extracted time-selected spectra from the EPIC-pn data.
The time selection is based on the OM exposures. Since each of the OM
exposures are only $\sim4\ks$ long, the corresponding time-selected X-ray
spectra extracted for the same time window as the OM exposures were
poor in signal-to-noise ratio. Therefore, we increased exposures of
each time-selected spectra to include the OM exposures and required
that exposures of two X-ray spectra from neighbouring time slices do
not overlap. The time-selection for the X-ray spectra is shown in
Fig. ~\ref{time_select_curve}.  The variations of the count rates in
$0.3-10\kev$, $0.3-2\kev$, $2-10\kev$ and OM U band are shown in
Fig.~\ref{time_sel_results}~(a). All the count rates are observed to be variable on $\sim 20$ks timescales
with fractional variability amplitudes \citep{2003MNRAS.345.1271V} of $F_{var} = 13.8\pm 2.8$,
$16.9\pm 5.0$ and $1.0 \pm 0.3$ for the hard, soft and UV fluxes, respectively (see Fig.~\ref{time_sel_results}). Since UV exposure is outside the X-ray flare for time interval C, the inclusion of X-ray flare may bias the correlation between X-ray and UV emission. We therefore used $45-60$ks interval of time slice C without X-ray flare to derive $F_{var}$ in both soft and hard band. We found that $F_{var}$ strengthens with similar values (hard band : $F_{var}=14.9\pm3.1\%$, soft band : $F_{var}=23.4\pm6.1\%$).

It is clear from Fig.~\ref{time_sel_results}~(a) and (b) that the 
the X-ray fluxes (and X-ray fluxes without C flare) in the soft and hard bands are correlated, 
The Spearman rank correlation is $0.89 ~(0.94)$ with a probability
that the two are not correlated $< 0.02 ~(0.01)$. However no such correlation
exists between the UV flux and the X-ray ones with a Spearman rank probability
that they are not correlated  being $< 0.2~(0.2)$ and $< 0.28~(0.2)$ for the 
soft and hard bands, respectively. Since each time selected observation
did not have enough quality to fit a sophisticated model we used 
an empirical one namely a combination of multicolour blackbody disc emission
(DISKBB) to represent the soft excess and a power-law for the
hard continuum.  The results of the
time-selected spectroscopy are listed in Table~\ref{pldbb1} and
plotted in Fig.~\ref{time_sel_results}~(b). Again, we see that while
the soft excess and the hard power law fluxes are correlated, the UV 
flux is not correlated to either of them.

The EPIC-pn lightcurve of II~Zw~177 in the $0.3-10$\kev~band has shown 
strong X-ray flares (see intervals C and E in
Fig.~\ref{time_select_curve}). In order to study short-term spectral variability, 
we extracted flare spectra from the intervals $35-45\ks$ 
and $95-105\ks$  of 2012 data. 

 We then fitted  the individual flare spectra in the C and E intervals with the
POWERLAW+DISKBB model. The best-fit parameters of both flares are listed in Table~\ref{pldbb1}. We found the
powerlaw fluxes $1.0\pm0.1\times10^{-12}$\funit~ and $1.3\pm0.1
\times10^{-12}$\funit~in $2-10$\kev~band for C and E flares, respectively. We
also found the soft X-ray excess fluxes $2.3\pm0.5\times10^{-12}$\funit~ and
$3.3_{-0.4}^{+0.5}\times 10^{-12}$\funit~in the $0.3-2$\kev~band for C and E
flares, respectively. Thus, our fitting clearly suggests that both powerlaw
flux and soft X-ray flux were higher during the flare state compared to
nearby intervals (see Table~\ref{pldbb1}. This
could be caused due to the rapid changes intrinsic to the corona producing higher
powerlaw emission and hence higher soft X-ray excess emission within the frame
of reflection model.

\begin{figure*}
    \centering
    \includegraphics[scale=0.6]{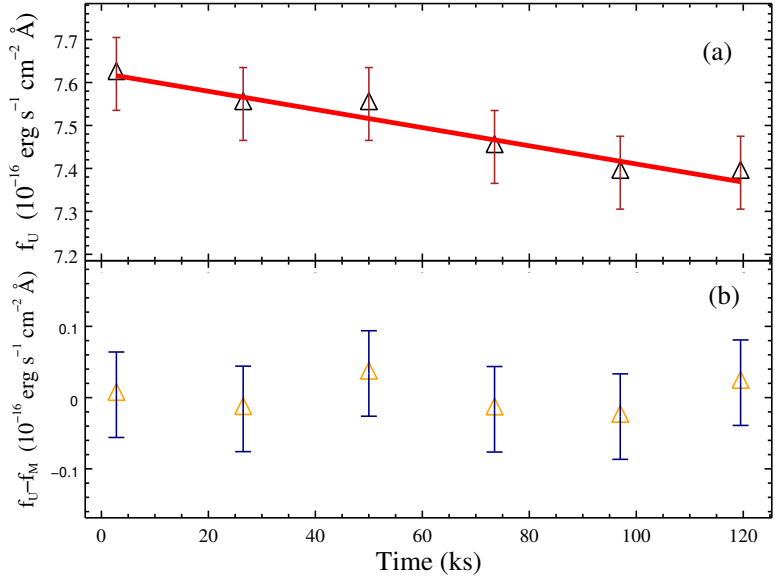}
    \caption{ The UV emission variability : (a) The best-fit linear model $f_{M}(t)=
    7.622-0.002~t$ to the observed UV emission $f_{U}$, showing gradual or secular decline. (b) The residuals $f_{U}-f_{M}$ show no significant variability on top of the secular decline.}
    \label{uv_variation}
  \end{figure*}

Furthermore, to study fractional variability amplitude ($F_{var}$) as a function of X-ray energy, we created rms spectrum using source lightcurves with a time binsize of 500s in the $0.3-10$\kev~band. We used the $0.3-0.5$, $0.5-1$, $1-2$ and $2-10$ \kev~bands to find $F_{var}$ for each energy band. We then plotted $F_{var}$ values together as shown in Fig.~\ref{rms_spec}. We also calculated the $F_{var}$ in the $0.3-2$\kev~in order to compare with $F_{var}$ in the $2-10$\kev~band using full EPIC-pn lightcurve with 500s bins. We found $F_{var}=20.1\pm0.4\%$ in the $0.3-2$\kev~band which is comparable to $F_{var}=20.1\pm1.9\%$ in the $2-10$\kev~band (see Fig.~\ref{rms_spec}). On the other hand, to examine the UV emission variability, we fitted a linear model to the UV lightcurve and found the best-fit function $P(t)= 7.622-0.002~t$, where $t$ the elapsed time in unit of ks. The fit resulted into $\chi^{2}/dof =0.39/4$. The best-fit model and data are shown in the upper panel of Fig.~\ref{uv_variation}. We then subtracted the best-fit predicted UV flux of the line model from the observed UV flux to derive the residual lightcurve. The UV lightcurve corrected for secular decline shows no significant variability (see lower panel of Fig.~\ref{uv_variation}). 

\section{Summary \& Discussion }

We studied the broadband UV to X-ray emission from II~Zw~177 using two
\xmm{} observations. Our main results are as follows.

$\it(i)$  Both observations show soft X-ray excess
emission below $2\kev$ which can be equally well modelled 
by a  blurred reflection and an intrinsic disc Comptonization
component. The ionized partially covering model does not describe
the data well.

$\it(ii)$ Between the two observations the flux in the $0.3$-$2$\kev~
varied by around 16\% while the flux in the $2$-$10$ \kev~ did not vary
much. For the blurred reflection model, the emissivity index
	did not vary and the height of the corona remains consistent ($2012 : h=
 3.3_{-1.4}^{+0.4}r_{g} , 2001: h=<4.7r_{g}$). Perhaps X-ray
 reprocessing is not contributing to the soft X-ray flux variation. 
 The small change in the soft band flux is likely due to the change
 in the shape of the powerlaw which appears to be steeper when the soft band flux was higher in 2001.

$\it(iii)$  For the long observation of 2012, the X-ray flux was found to be
variable on timescale of $\sim20$ ks with fractional variability
amplitude of $F_{var} \sim 15$\%. Time resolved spectral
fitting also revealed that the soft excess represented by  
thermal emission is well correlated with the hard powerlaw flux.

$\it(iv)$  Extension of the best fit X-ray thermal Comptonization model
shows that the observed UV emission is significantly higher than that
expected from a standard accretion disc. This discrepancy can be removed
if the black hole mass is $\sim 4 \times 10^7 M_\odot$ which is significantly
larger than $2 \times 10^6 M_\odot$ reported in the literature.

$\it(v) $ The UV flux is found to be weakly variable at a level of $F_{var} \sim 1$\%. However, the UV flux
is not correlated to the total X-ray flux nor to the X-ray fluxes
in the  $0.3$-$2$ and $2$-$10$ \kev~bands. The UV flux is also not correlated
to the soft excess or powerlaw X-ray components. The gradual decline of the observed UV
emission such much longer variability timescale compared to $\sim20$ ks, and may be related 
to the secular decline resulting from changes in  the accretion rate.

Within the context of the blurred reflection model our results indicate that
the geometry i.e., location of the hard X-ray producing corona does not vary
in this source on long timescale. The soft X-ray excess and powerlaw fluxes
are similar and other spectral components are also found consistent except
slight variation in photon index. Hence the marginal variability
is not caused by the X-ray reflection on long timescale. The
photon index tends to be more steeper in 2001 observation. This
can be seen clearly in the results of the Comptonization model (see Table-2).

The Comptonization model also explains full X-ray band including the soft excess
equally well to both 2012 and 2001 observations. Despite the constant temperature of cool
Comptonization medium, the slight change in accretion rate ($2012~\rm observation : L/L_{E} = 0.65\pm0.01$,
$2001~\rm observation : L/L_{E} = 0.8_{-0.1}^{+0.8}$) and hence the change in gravitational
energy may be causing the long-term variation in high energy emission and hence photon
index.

The time-resolved
spectroscopy of 2012 observation shows that powerlaw slope was found similar in
all time intervals. In particular, we did not observe a clear correlation between the photon index and the powerlaw flux. 
Hence, the powerlaw flux variability is unlikely to be caused by the variations in the seed flux from accretion disc.
The variability is most likely intrinsic to the hot corona.   If the
hard X-ray source is compact and close to the black hole, then strong light
bending effects make the powerlaw component strongly variable and soft X-ray
excess weakly variable \citep{2004MNRAS.349.1435M}. This is in contrast to the
variability seen in the soft X-ray excess (i.e., the reflection component) and
powerlaw components. The soft excess (i.e., the
reflection component) and the powerlaw fluxes are correlated with similar
fractional variability amplitudes ($F_{var}\sim15\%$). Moreover, this is further strongly supported by the rms spectra shown in Fig.~\ref{rms_spec} suggesting similar variability in soft and hard bands. 
Hence, intrinsic variations in the corona, possibly unrelated 
to the changes in the height, must be responsible for the
observed variability of the powerlaw and its correlation with the soft X-ray excess.
In the blurred reflection model, the soft X-ray excess is expected to follow the intrinsic variations 
in the coronal emission.

The observed UV emission
is weakly variable on timescale of $\sim20\ks$  with $F_{var}\sim1\%$ and
further shows no correlation with X-ray emission in the soft ($0.3-2\kev$) and hard ($2-10\kev$) bands or X-ray spectral
components. If the UV emission is the thermal emission from a standard disc, the U band flux with
a central wavelength of $3440$\text{~\AA} must arise from radii near $\sim
  2000r_g$ in the disc where the temperature is
  $\sim10^{4}{\rm~K}$ for $M_{BH}\sim
  2\times10^{6}{\rm~M_{\odot}}$ and accretion rate $\dot{m}\sim 0.7$
  relative to the Eddington rate (as suggested by the OPTXAGNF model). At these radii, the viscous timescale 
 is very long ($t_{vis} \sim 30-3000{\rm~years}$ for $r/H \sim 10-100$ and $M_{BH}\sim
 2\times10^6{\rm~M_{\odot}}$). The light crossing time is the time delay between the X-ray emission from a compact corona around the SMBH and the UV emission from the disc. If the UV emission is caused by the reprocessing of X-rays into the disc, then the UV emission is believed to lag behind the X-rays by the light crossing time. If the variable X-rays is dominated by the variations in the seed UV photons, then the X-ray emission is expected to lag behind the UV emission again by the light crossing time. Assuming only viscous heating of accretion disc, the disc emission from each radius can be described as a blackbody with distinct temperature. According to Wein's law, the blackbody temperature can be deduced from the effective wavelength ($\lambda_{eff}$) peaking in the observed bandpass. Comparing this temperature as from the standard accretion disc, one can find the radius of accretion disc for observed emission in a given bandpass (e.g., \citealt{2007MNRAS.375.1479S}). Thus, the light crossing time between the emitting regions of X-ray and UV emission can be derived to be 
\begin{equation}
\tau_{cross} \approx 2.6\times10^5 \left(\frac{\lambda_{eff}}{3000~\text{\AA}}\right)^{4/3}   \left(\frac{\dot{M}}{\dot{M}_{Edd}}\right)^{1/3} \left(\frac{M_{BH}}{10^8M_{\odot}}\right)^{2/3}{\rm~sec}.
\end{equation}

Where $\left(\frac{\dot{M}}{\dot{M}_{Edd}}\right)$ is the relative accretion rate with respect to Eddington rate and $M_{BH}$  is the black hole mass of an AGN.  For II~Zw~177, $\tau_{cross}\sim20\ks$ for $\lambda_{eff}=3440\AA$, $\frac{\dot{M}}{\dot{M}_{Edd}}=0.7$, $M_{BH}=2\times10^{6} M_\odot$.  The absence of UV/X-ray correlation on $20\ks$ time scale clearly implies that the observed UV emission is not dominated by the reprocessed emission.

  The  extremely weak ($F_{var}\sim1\%$) variability of the UV with gradual  decline could be part
  of large amplitude variability on very long time scale such as the viscous time scale. If this is case, any fluctuations
  in the UV emission on top of the secular decline due  to changes in the accretion rate, may be related to the variations 
  in the reprocessed emission. In Fig.~\ref{uv_variation}, we show the residuals obtained by subtracting a linear model representing secular decline from the observed UV emission. There are no significant variation within these residuals that would be attributed to low level X-ray reprocessing. Thus, we find no clear evidence for reprocessed UV emission from II~Zw~177. High S/N data are required to fully address this possibility.

In the intrinsic disc Comptonized (i.e., the OPTXAGNF) model, the soft X-ray excess arises from the inner disk itself.
Therefore, the variations in the UV and the soft excess emission are expected to be correlated. The absence of any correlation between the UV and soft excess
suggests that the soft excess is not dominated by the thermal Comptonization in the disc. As discussed earlier, the correlation 
between the soft excess and powerlaw emission are consistent with that expected in the blurred reflection model provided that
the variations in the powerlaw component are intrinsic to the corona rather than due to the changes in the height of the corona.

Alternatively, perhaps the UV emission is intrinsic to the disc and the variability is due to
the variation in some absorbing material in the line of sight. A varying
covering fraction at a few percentage level may lead to the $1$\% variability
seen. However, it is difficult (but perhaps not impossible) to see how a patchy absorbing region extending
to $\sim 2000 r_g$ could vary on such short timescale as $\sim$ 20 ks. Such a model would also
require a larger black hole mass for the system which would aggravate the problem of the short timescale
variability of the absorbing clouds.

Essentially, our results indicate that the short term variability of the X-ray
and UV emission of AGNs can be complex but have the potential to provide 
important clues regarding the nature and geometry of the central source. There
is evidence that the geometry of the central regions are different for
different AGNs and even for a particular AGN, it may change in timescale of
years. A comprehensive observational programme for studying simultaneous UV and X-ray
emission from several AGNs over different timescale may provide a much clearer picture
of the dynamic nature of these sources. Such an endeavour may be possible by the future
X-ray mission {\it Astrosat} \citep{2014SPIE.9144E..1SS} which will have the capability of sensitively measuring the
simultaneous X-ray and UV emission from AGNs.

\section{Acknowledgement}
 An anonymous referee is gratefully acknowledged for the suggestions/comments which improved the paper.
MP and PKP gratefully acknowledge support by the CSIR, New Delhi under the NET fellowship program.
This research has made use of ($i$) NASA's Astrophysics Data System,
($ii$) the SIMBAD database, operated at CDS, Strasbourg, France,
($iii$) observations obtained with XMM-Newton, an ESA science mission
with instruments and contributions directly funded by ESA Member
States and NASA, and ($iv$) data, software and/or web tools obtained
from the High Energy Astrophysics Science Archive Research Center
(HEASARC), a service of the Astrophysics Science Division at NASA/GSFC
and of the Smithsonian Astrophysical Observatory's High Energy
Astrophysics Division.

\label{lastpage}

\newcommand{\pasp}{PASP} \def\apj{ApJ} \def\mnras{MNRAS}
\def\aap{A\&A} \def\apjl{ApJ} \def\aj{aj} \def\physrep{PhR}
\def\pre{PhRvE} \def\apjs{ApJS} \def\pasa{PASA} \def\pasj{PASJ}
\def\nat{Nat} \def\ssr{SSRv} \def\aapr{AAPR} \def\araa{ARAA}

\bibliographystyle{mn2e} 
\bibliography{refs}

\end{document}